\documentclass[12pt]{article}
\pdfoutput=1
\usepackage[nosort]{cite}
\usepackage[normalem]{ulem}
\usepackage{lipsum,calligra,pbsi}
\usepackage{CJK}

\usepackage{diagbox}  

\usepackage[dvipsnames]{xcolor}
\usepackage{epsfig}
\usepackage{amsfonts}
\usepackage{amscd}
\usepackage{latexsym}
\usepackage{amsmath, amssymb, amsthm}
\usepackage{verbatim}
\usepackage{color}
\usepackage{fancyhdr}
\usepackage{hyperref}
\usepackage{tikz}
\usepackage{slashed}
\usepackage{multirow}
\usepackage{braket}
\usetikzlibrary{calc}
\usetikzlibrary{topaths}
\usetikzlibrary{decorations}
\usetikzlibrary{decorations.pathmorphing}
\usetikzlibrary{arrows,decorations.markings,cd}
\usetikzlibrary{calc,arrows,cd,decorations.markings, snakes}
\tikzset{
->-/.style args={#1rotate#2}{decoration={markings, mark=at position #1 with {\arrow[scale=1.5,rotate = #2 ]{stealth}}}, postaction={decorate}}
}
\usetikzlibrary{shapes.geometric}
\usetikzlibrary{knots}
\usepackage{tikz-3dplot}

\usepackage{float}
\usepackage{draft}
\usepackage{graphicx, subfig}
\usepackage{mciteplus}
\usepackage{skak}
\usepackage{bbm}

\usepackage{cleveref}
\usepackage{algorithm}
\usepackage[letterpaper,left=1.1in,top=1in,right=1.1in,bottom=1.3in]{geometry}
\usepackage{makecell}
\usepackage{enumerate}
\usepackage{tablefootnote}
\usepackage{longtable}
\usepackage{mathrsfs}

\numberwithin{equation}{section}

\usepackage[us,hhmmss]{datetime}

\usepackage{pifont}

\usepackage[scr=euler]{mathalfa}

\usepackage[english]{babel}
\usepackage{soul}
\usepackage{upgreek}
\usepackage[scr]{rsfso}

\DeclareMathAlphabet{\zapfcal}{OT1}{pzc}{m}{it}

\DeclareFontFamily{OT1}{pzc}{}
\DeclareFontShape{OT1}{pzc}{m}{it}{<-> s * [1.10] pzcmi7t}{}
\DeclareMathAlphabet{\mathpzc}{OT1}{pzc}{m}{it}

\colorlet{mylinkcolor}{NavyBlue}
\colorlet{mycitecolor}{Aquamarine}
\colorlet{myurlcolor}{Aquamarine}
\newcommand\myshade{90}

\hypersetup{
  linkcolor  = mylinkcolor!\myshade!black,
  citecolor  = mycitecolor!\myshade!black,
  urlcolor   = myurlcolor!\myshade!black,
  colorlinks = true,
}

\def\U{\mathrm{U}(1)}

\def\cA{\mathcal{A}}

\def\bZ{\mathbb{Z}}
\def\bR{\mathbb{R}}
\def\cT{\mathcal{T}}

\begin{document}

\begin{titlepage}

\title{Peeling off Abelian Anyons and Proliferation Transitions}

\begin{CJK*}{UTF8}{gbsn}
\author{Meng Cheng (程蒙)$^{1,2}$ and Nathan Seiberg$^2$}
\end{CJK*}
 \address{${}^1$ Department of Physics, Yale University, New Haven, Connecticut 06511, USA}
 \address{${}^2$ School of Natural Sciences, Institute for Advanced Study, Princeton, NJ}

\abstract{
\noindent
We study the invertible one-form global symmetry of a (2+1)-dimensional topological quantum field theory (TQFT).  A universal description of this symmetry  ``peels off'' all the Abelian anyons from the rest of the TQFT.   This allows us to study this symmetry and some of its consequences, independently of the details of the rest of the system. We then describe these Abelian anyons using $\U$ gauge fields with Chern-Simons terms and couple them to a scalar field, thus constructing a particular non-topological theory.  This new theory, our transition field theory, exhibits a phase transition from the original TQFT to another phase.  The latter is either gapless or another TQFT.  This discussion generalizes and extends our earlier work \cite{Cheng:2026qax}.  In particular, we give a more detailed description of the global symmetries and their fates in the various phases.}
\end{titlepage}

\tableofcontents

\section{Introduction}

In (2+1)-dimensional topological quantum field theories (TQFT),\footnote{Throughout this work, we limit ourselves to bosonic theories.  The extension to fermionic or electronic theories is straightforward.} Abelian anyons play a distinguished role as they generate (invertible) one-form symmetries of the theory \cite{Gaiotto:2014kfa, Hsin:2018vcg}. Among other things, they underlie the structure of zero-form symmetry enrichment \cite{barkeshli_symmetry_2019}.

When the entire TQFT is Abelian \cite{Belov:2005ze, Kapustin:2010hk}, it is convenient to describe the theory as a Chern-Simons theory of $\U$ gauge fields. This is to be contrasted with non-Abelian TQFTs, most of which can be described by Chern-Simons gauge theories with non-Abelian gauge fields \cite{Witten:1988hf, Moore:1989yh}.  These theories are more involved to work with. It is thus desirable to have a presentation of the TQFT, where the Abelian anyons are described by $\U$ gauge fields, even when the full TQFT is non-Abelian. Such a presentation was developed in \cite{Cordova:2017vab,  Hsin:2018vcg, Cheng:2026qax}, where the focus was on a single $\bZ_n$ group of Abelian anyons.

This formalism was used for various purposes.  In particular, \cite{Cheng:2026qax} presented a non-topological field theory that exhibits a transition out of a TQFT $\cal T$, driven by the proliferation of an Abelian anyon in $\cT$, where a single complex scalar field captures the dynamics of the anyon.

In this work, we generalize the description of the TQFT in \cite{Cordova:2017vab,  Hsin:2018vcg, Cheng:2026qax}, and present all the Abelian anyons in the TQFT using $\U$ gauge fields with Chern-Simons terms.

As an application of this description of the TQFT, we revisit the transition field theory studied in \cite{Cheng:2026qax}.  It was shown in \cite{Cheng:2026qax} that once the anyon type is fixed, the independent data of the theory is a single integer $p$.  Our more general form of the theory allows us to reproduce the result of \cite{Cheng:2026qax}, while keeping track of all other Abelian anyons in the theory.  This leads us to a more detailed description of the theory's global symmetry and phases.

In particular, one version of the transition theory has an exact $\U_{\cal B}$ global symmetry, and we obtain useful new expressions for its current and the minimal charge operators.  Also, the analysis of the $\U_{\cal B}$ current leads us to the most general coupling of the theory to a background $\U_{\cal B}$ gauge field.

In another version of the theory, this $\U_{\cal B}$ symmetry is explicitly broken to a discrete $\bZ_\ell \subset\U_{\cal B}$ global symmetry.  But this $\bZ_\ell$ symmetry cannot be further explicitly broken, because even though it is a faithful symmetry of the theory, it does not act on local operators.\footnote{Throughout this paper, we distinguish between symmetries that act faithfully, with nontrivial symmetry operators, and symmetries that act only via enrichment.  See e.g., the discussion in \cite{Seiberg:2025bqy}.}

We then provide a deeper understanding of the Higgs phase of the $p=0$ theory.  Here, the version of the $p=0$ theory with $\U_{\cal B}$ is gapless.  $\U_{\cal B}$ is spontaneously broken to its $\bZ_\ell$ subgroup \cite{Cheng:2026qax}, and hence, the low-energy theory is similar to that of an ordinary superfluid.  It has stable vortices, which are charged under a certain exact global one-form symmetry.  This one-form symmetry is unbroken, and relatedly, these vortices are the boundaries of logarithmically confining strings.  In the version of the theory where this $\U_{\cal B}$ global symmetry is explicitly broken to its $\bZ_\ell$ subgroup, these vortices still exist.  And they are still charged under the one-form global symmetry, and hence, they still bound confining strings.  But unlike the case with the continuous $\U_{\cal B}$ symmetry, now the strings have nonzero tension and the related line operators exhibit an area law.

In Section~\ref{Abelian anyons}, we discuss the Abelian anyons $\cA$ in the TQFT $\cT$ and the invertible one-form symmetry generated by $\cA$. We show that the Hilbert space on a closed surface with a single $s\in \cT$ insertion is empty (i.e., zero-dimensional) for all surfaces, if and only if $s$ transforms nontrivially under $\cA$.  This result is important in the later sections.

In Section~\ref{AintermsofU1}, we provide a presentation of the TQFT $\cT$, such that all the Abelian anyons are described explicitly using $\U$ gauge fields, generalizing the duality in \cite{Cordova:2017vab, Hsin:2018vcg, Cheng:2026qax}.

In Section~\ref{proliferationtransitions}, we revisit the transition theory studied in \cite{Cheng:2026qax}, where $\cT$ is coupled to a complex scalar. We make sure that the presentation of $\cal T$ in Section~\ref{AintermsofU1} includes all the discrete data of the transition theory.  Some technical details of this analysis are presented in Appendix \ref{KvshatK}.

Section~\ref{globalU1} analyzes the global symmetry of the transition theory.

In Section~\ref{sec:phases}, we study the phases of the transition theory, generalizing \cite{Cheng:2026qax}, and providing additional results.  In particular, the better understanding of the global zero-form and one-form symmetries and the associated vortices leads to a more complete understanding of the relation between the phase transition of the $p=0$ theory and gauging a one-form symmetry of $\cal T$ (aka anyon condensation).

In Section~\ref{sec:coupling to external A}, we discuss the coupling of the theory to an external $\U_A$ gauge field, and clarify the relation between this background gauge field, the global $\U_{\cal B}$ symmetry of Section \ref{globalU1}, and the background gauge field for that symmetry, discussed in Section \ref{BackgroundB}.

Finally, in Section~\ref{conclusions}, we summarize our results and highlight the crucial role the problem's global symmetries play in our analysis.

\section{Abelian anyons $\cal A$ in a TQFT}\label{Abelian anyons}

\subsection{Characterizing $\cal A$ and our notation}\label{Abeliannotation}

We start with the discussion of the Abelian anyons ${\cal A}\subseteq{\cal T}$ in the TQFT $\cal T$.  They generate the invertible one-form symmetry of the TQFT.

Importantly, $\cal A$ does not have to be modular.  More technically, it is referred to as pre-modular.  It describes the one-form symmetry\footnote{We distinguish between the Abelian TQFT $\cal A$ and the one-form symmetry ${\cal G}_{\cal A}^{(1)}$.} ${\cal G}_{\cal A}^{(1)}=\prod_i \bZ_{n_i}^{(1)}$, generated by $a_1, a_2, \cdots, a_N$, satisfying $a_i^{n_i}={\bf 1}$.\footnote{Here, we made a choice of basis.  In general, $\cal G_A$ can be written as a product of cyclic groups in several different ways, and we can pick different generators for each factor.} The spins of these anyons and their braidings are
\ie\label{braidele}
&h(a_i)=\frac{p_i}{2n_i}\bmod 1\\
&B(a_i,a_j)=\frac{p_{ij}}{\gcd(n_i,n_j)}\bmod 1\,,\\
&p_{ij}\in \bZ\qquad, \qquad p_{ii}=p_i\qquad, \qquad p_in_i\in 2\bZ\,.
\fe
At this level, $p_{ij}$ is meaningful only modulo $\gcd (n_i,n_j)$, except $p_{ii}=p_i$, which is meaningful modulo $2n_i$.  These integers characterize the anomaly in ${\cal G_A}^{(1)}$ \cite{Gaiotto:2014kfa, Hsin:2018vcg}.

We will find it convenient to use the following notation.  The anyon $s=\prod_{i=1}^N a_i^{s_i}\in {\cal A}$ will be characterized by the $N$-dimensional vector $s$.\footnote{We denote by $s$ both the anyon and the corresponding vector in our basis.  Also, we use the basis of anyons $a_i$.  This is not the $i$-th component of an anyon $a$.  We hope that this will not cause confusion.}  We will also use the following $N\times N$ dimensional matrices
\ie\label{BMDdef}
    &D_{ij}=n_i \delta_{ij}\,,\\
&B^{ij}={p_{ij}\over\gcd(n_i,n_j)}\,.
\fe
Here, $p_{ij}$ are integers rather than integers modulo $\gcd(n_i,n_j)$, and we recognize the elements of the matrix $B^{ij}$ as the braiding \eqref{braidele} of $a_i$ and $a_j$.

The freedom in shifting $p_{ij}$ by an integer multiple of $\gcd(n_i,n_j)$ for $i\ne j$ and by an integer multiple of $2n_i$ for $i=j$ can be written as
\ie\label{Bshift}
B\to B-{\bf L}-{\bf L}^\mathrm{T} \qquad, \qquad {\bf L}\in \bZ^{N\times N}\,.
\fe
Similarly, the vector $s$  can be shifted by
\ie\label{sshift}
s\to s-D{\bf k} \qquad, \qquad {\bf k} \in \bZ^N\,.
\fe
These shifts do not affect the invariant information in $\cal A$ and in the choice of anyon type $s\in {\cal A}$.

The data \eqref{BMDdef} can be summarized as follows.  The theory has a ${\cal G_A}^{(1)}=\prod_i \bZ_{n_i}^{(1)}$ global symmetry.  We can couple it to $N$ background gauge fields ${\frak B}^i$.  These are $\bZ_{n_i}$ 2-cocycles.  The anomaly means that the (2+1)-dimensional theory is not invariant under gauge transformations for these fields.  This can be described by a (3+1)-dimensional invertible anomaly theory
\ie\label{anomalytheory}
\sum_i {2\pi p_i\over 2 n_i}{\cal P}({\frak B}^i) +\sum_{i\ne j}{\pi p_{ij}\over \gcd(n_i, n_j)} {\frak B}^i\cup {\frak B}^j\,,
\fe
where ${\cal P}({\frak B}^i) $ is the Pontryagin square.  (As a check, this expression changes by $2\pi\bZ$ when the cocycle ${\frak B}^i$ is shifted by $n_i$.)

\subsection{The invertible one-form symmetry -- the action of $\cal A$ on $\cal T$}\label{AonT}

Every anyon $s\in {\cal T}$ transforms under $\prod_i \bZ_{n_i}^{(1)}$ as $\exp \left({2\pi i n_s^i\over n_i}\right)$.  For the Abelian anyons $s\in {\cal A\subseteq T}$, we can use our matrix notation
\ie
&n_{s}^i=n_i\left(B s\right)^i \bmod n_i\\
&p_{ij}={\gcd(n_i,n_j)n_{a_i}^j\over n_j}\bmod (\gcd(n_i,n_j))={\gcd(n_i,n_j)n_{a_j}^i\over n_i}\bmod(\gcd(n_i,n_j))\\
&B(s,s')=s^\mathrm{T}Bs'\bmod 1\,.
\fe

The invertible one-form symmetry $\cal A$ has many important applications, as it constrains the correlation functions.  In particular, we will use the following result later.

Let $\Sigma_g$ be a Riemann surface of genus $g$, and denote by $\mathcal{H}_s(\Sigma_g)$ the Hilbert space on $\Sigma_g$ with a single $s\in {\cal T}$ insertion. Then, if $s$ transforms nontrivially under $\cal A$, i.e., $n_s^i$ is nonzero for at least one value of $i$, $\mathcal{H}_{s}$ is zero-dimensional.  To see that, wrap the anyon $a_i $ around $s$ to find a phase, and then unwrap it by pulling it around $\Sigma_g$ to conclude that the correlation function is zero.

More interestingly, the converse statement is also true.  If $s$ transforms trivially under $\cal A$, i.e., $n^i_s=0$ for all $i$, then for sufficiently large $g$, we have
\begin{equation}\label{dimHx}
    {\rm dim}\,\mathcal{H}_{s}(\Sigma_g)>0.
\end{equation}

To see this\footnote{We thank GPT 5.5 for an inspiring ``conversation'' about this argument.}, we use the Verlinde formula for the dimension
\begin{equation}
     {\rm dim}\,\mathcal{H}_{s}(\Sigma_g) = \sum_{x\in \cT}S_{sx}S_{1x}^{1-2g} = \sum_{a\in \cA} S_{sa}S_{1a}^{1-2g}+\sum_{x\notin \cA}S_{sx}S_{1x}^{1-2g}.
\end{equation}
Since $s$ braids trivially with all the elements in $\cal A$, $S_{sa}=S_{s1}=\frac{d_s}{\cal D}$.  (Here, $d_x$ is the quantum dimension of $x$ and ${\cal D}={1\over S_{11}} =\sqrt{\sum_{x\in {\cal T}}d_x^2}$ is the total dimension.) Using also $S_{1x}=\frac{d_x}{\cal D}$, the dimension of the Hilbert space can be written as
\begin{equation}
     {\rm dim}\,\mathcal{H}_{s}(\Sigma_g) = {{\cal D}^{2g-2}}\left( |\cA|d_s+{\cal D}\sum_{x\notin \cA}S_{sx}d_x^{1-2g}\right).
\end{equation}
Since for $x\notin {\cal A}$, $d_x>1$, for large enough $g$, the absolute value of the second term is arbitrarily small, while the first term is a positive constant independent of $g$. Therefore, the dimension must be a positive integer, i.e., \eqref{dimHx}. In fact, the argument also gives the asymptotic value of the dimension to be ${\cal D}^{2g-2}d_s|\cA|$ for $g\gg 1$.

Let us illustrate it with two  examples:
\begin{itemize}
    \item $\cT$ is Abelian, so $\cal A=T$, and it is modular.  Therefore, there is no $s\ne 1$ that braids trivially with $\cal A$.  Consequently, the one-point function for $s\ne 1$ vanishes for every genus $g$.
\item $\cT={\rm SU}(2)_k$, and $\cA=\{1, j=k/2\}$.  The anyons with half-integer $j$ transform nontrivially under $\cal A$ and indeed, their one-point functions vanish.  Anyons with integer $j$ braid trivially with $\cal A$.  Indeed, they have nonzero one-point functions on every surface with genus $g\ge 1$. Specifically, an anyon with $j\in \bZ$ has a nonzero torus one-point function with the line with $j\over 2$ running around the torus.
\end{itemize}

\subsection{Focusing on a particular anyon $q \in {\cal A}$}\label{focusq}

In Section \ref{proliferationtransitions}, we will introduce a complex scalar field $\Phi$ corresponding to a particular Abelian anyon.  In preparation for this discussion, we study here some aspects of $\cal A$ that depend on a choice of anyon $q \in {\cal A}$.

The anyon $q$ can be expressed as $q=\prod_{i=1}^Na_i^{q_i}$, and we denote it by the vector $q$ with components $q_i$.\footnote{ We hope that using $q$ both for an anyon $q\in {\cal A}$ and for the $N$-dimensional vector of integers that represent it in some basis of generators of $\cal A$ will not cause confusion.}  Its order $n$ is the smallest  positive integer such that
\ie\label{defbe}
nD^{-1}q\in \bZ^N \,
\fe
with $D$ of \eqref{BMDdef}. This means that the anyon $q^n$ is the identity.  Hence, $q$ generates ${\cal A}_q\subseteq {\cal A}$ with the corresponding one-form symmetry group ${\cal G}_{{\cal A}_q}^{(1)} =\bZ_n$.

Using \eqref{BMDdef}, we also have\footnote{The vector $q$ can be shifted as $q\to q-D{\bf k}$ with ${\bf k}\in \bZ^N$ without changing the anyon. (See \eqref{sshift}.) This transformation does not change $n$, but it shifts $u\to u -nBD{\bf k}$ and $p=q^\mathrm{T} u\to p -n{\bf k}^\mathrm{T}DBq - nq^\mathrm{T}BD{\bf k} +n{\bf k}^\mathrm{T} DBD{\bf k} $.  Importantly, it does not shift $p\bmod 2n$. \label{shiftqT}}
\ie\label{uofqd}
&u=n B q \in \bZ^N\\
&p=q^{\mathrm T}u \bmod 2n\,.
\fe
As a check, $h(q)={q^\mathrm{T}Bq\over 2}\bmod 1={p\over 2n}\bmod 1$, which is consistent with the fact that $p\bmod 2n$ is the anomaly in the $\bZ_n$ symmetry generated by $q$.

A consequence of the anomaly $p$ is that all the anyons in ${\cal A}_q$ transform under $\bZ_n$ as powers of $e^{2\pi i p\over n}$.  These are powers of $e^{2\pi i L\over n}$  with
\ie
L=\gcd(n,p)\,.
\fe
Hence, ${\cal A}_q$ realizes faithfully only $\bZ_{n\over L}={\bZ_n\over\bZ_L}$.

Let us examine how the other anyons in $\cal A$ transform under $\bZ_n$.

Every anyon $s\in {\cal A}$ braids with $q$ as $e^{2\pi i n_s\over n}$ with 
\ie
n_s=s^\mathrm{T} u\,.
\fe
Let $ \ell$ be the greatest common divisor of $n$ and all $n_s$,
\ie\label{elldefu}
\ell = \gcd(n,u^1,u^2,\cdots,u^N)\,.
\fe
Using that, the anyon $q^{n \over \ell}$ braids trivially with all the anyons in $\cal A$.   Note, however, that for $\ell\ne 1$, it braids nontrivially with other anyons in $\cal T$.

Another consequence of \eqref{elldefu} is that there is a vector of integers $\tau$ such that $\tau^\mathrm{T}u=\ell \bmod n$.  Correspondingly, there is an anyon $\tau\in {\cal A}$ with
\ie\label{xuell}
n_\tau = n\tau^\mathrm{T}B q=\tau^\mathrm{T}u=\ell\bmod n\,.
\fe
Then, every anyon $s\in \cal A$ can be written as $\tau^{n_s\over \ell}$ times an anyon that braids trivially with $q$. This means that only $\bZ_{n\over \ell} = {\bZ_n\over \bZ_\ell}$ is realized faithfully in $\cal A$.

We learn that $L$ describes how the group ${\cal G}_{{\cal A}_q}^{(1)}=\bZ_n$ acts on ${\cal A}_q$, and $\ell$ describes how it acts on $\cal A$.

Another interpretation of this $\bZ_{n\over \ell} = {\bZ_n\over \bZ_\ell}$ quotient is the following.  Let $\cA_q^\perp \subseteq \cal{A}$ be the anyons in $\cal A$ that braid trivially with $q$ (and therefore with ${\cal A}_q$)
\ie\label{Aperpd}
\cA_q^\perp=\{r\in \cA| r^\mathrm{T}u=0\bmod n\}={\{r\in \bZ^N|r^\mathrm{T}u=0\bmod n\}\over (r \sim r -D{\bf k} )}\qquad , \qquad {\bf k}\in \bZ^N\,.
\fe
Since every anyon $s\in \cal A$ can be written as $\tau^{n_s\over \ell}$ times an anyon in $\cA_q^\perp$, and since $\tau^{n\over \ell}$ braids trivially with $q$, i.e., $\tau^{n\over \ell}\in\cA_q^\perp$,\footnote{We should point out, however, that as an element in $\cal A$, the anyon $\tau^{n\over \ell}$ could be nontrivial.  Also, note that this is a standard quotient of groups.  Later, we will use a similar notation for gauging a one-form symmetry.}
\ie\label{Aquotient}
{{\cal G_ A}\over {\cal G}_{\cA_q^\perp}} \cong \bZ_{n\over \ell}\,.
\fe
This quotient means that if $\tau$ satisfies \eqref{xuell}
\ie\label{tauplusr}
\tau^\mathrm{T}u=\ell\bmod n\,,
\fe
so does $\tau+r$ with $r\in
\cA_q^\perp$ of \eqref{Aperpd}.

Finally, in the special case $\ell=n$, all the anyons in $\cal A$ braid trivially with $q$ and hence ${\cal A} = {\cal A}_q^\perp$.  Then, $\tau$ above is trivial and can be set to zero.

Following the discussion in Section \ref{AonT}, there exists a surface $\Sigma$, perhaps of high genus, such that
\ie\label{nonzeroRie}
\langle q^{n\over \ell}\rangle_\Sigma \ne 0 \qquad, \qquad \ell\ne 1\,.
\fe

 To summarize, we picked $q\in {\cal A}\subseteq {\cal T}$.  Then:
\begin{itemize}
    \item[$q^{n\over L}$] braids trivially with all the anyons in ${\cal A}_q\subseteq {\cal A}\subseteq {\cal T}$.  It generates
    \ie\label{ZsubL}
{\cal A}_{q^{n\over L}}\cong {\cal A}_q\cap {\cal A}_q^\perp  \qquad , \qquad L=\gcd(n,p)\,.
\fe
In other words, only ${\cal G}_{{\cal A}_{q^{n\over L}}}^{(1)} =\bZ_L^{(1)}\subseteq \bZ_n^{(1)}$, generated by $q^{n\over L}$ is included in ${\cal G}_{{\cal A}_q^\perp}^{(1)}$.
\item[$q^{n\over \ell}$] braids trivially with all the anyons in ${\cal A}\subseteq {\cal T}$.  It generates
\ie\label{Zsubell}
{\cal A}_{q^{n\over \ell}}\subseteq {\cal A}_{q^{n\over L}} \qquad,\qquad \ell = \gcd(n,u^1,u^2,\cdots,u^N) \,.
\fe
In other words, ${\cal G}_{{\cal A}_{q^{n\over \ell}}}^{(1)} =\bZ_\ell^{(1)}\subseteq \bZ_L^{(1)}\subseteq \bZ_n^{(1)}$, generated by $q^{n \over \ell} $ does not act on $\cal A$.
     \item[$q^n$] braids trivially with all the anyons in ${\cal T}$.  It is a trivial anyon.
\end{itemize}

Returning to the two examples at the end of Section \ref{AonT}:
\begin{itemize}
    \item $\cT$ is Abelian, so $\cal A=T$.    Hence, $\ell=1$ and \eqref{nonzeroRie} is not effective.  See Section 7.1 in \cite{Cheng:2026qax}.
\item ${\cal A}={\cal A}_q\subset {\cal T}$.  In this case $\ell=L$.  An example is SU(2)$_k$, where $L=\gcd(2,k)$ and then, \eqref{nonzeroRie} is nontrivial for even $k$.  See Section 7.2 in \cite{Cheng:2026qax}.
\end{itemize}

\section{Representing $\cA$ using $\U$ gauge fields}\label{AintermsofU1}

\subsection{$\cal K$}\label{KintermsofA}

Here, we extend the pre-modular theory $\cal A$ to a modular theory. For every pre-modular $\cal C$, there is a canonical procedure to do that leading to the Drinfeld center of $\cal C$.  In our case, of Abelian $\cal A$, we can make it very explicit. We construct an Abelian TQFT $\cal K$ based on the $2N\times 2N$-dimensional K-matrix:
\ie\label{KforcalK}
    K = \begin{pmatrix}
        -DBD && D \\
        D && 0
    \end{pmatrix}\,,
\fe
where $D$ and $B$ are $N\times N$ matrices given by \eqref{BMDdef}.
We denote the $\U^{2N}$ gauge fields as $\begin{pmatrix} b \\ c\end{pmatrix}$, where $b$ and $c$ are $N$-dimensional vectors, and we label the anyons as $\begin{pmatrix}\omega_b \\ \omega_c\end{pmatrix}$ with $\omega_b , \omega_c \in \bZ^N$.   Given the identifications
\ie\label{omegaident}
 \begin{pmatrix}\omega_b \\ \omega_c\end{pmatrix} \sim  \begin{pmatrix}\omega_b \\ \omega_c\end{pmatrix}+ K{\bf l} \qquad, \qquad {\bf l}\in \bZ^{2N}\,,
\fe
the $i$-th entry in $\omega_b$ is identified modulo $n_i$ and similarly for $\omega_c$.

The form of the K-matrix \eqref{KforcalK} for $\cal K$ shows that $\cal K$ is a $\cal G_A$ gauge theory with a Dijkgraaf-Witten twist \cite{Dijkgraaf:1989pz} given by $-DBD$.

The matrix $K$ can be transformed by ${\rm GL}(2N,\bZ)$ transformations without affecting the TQFT $\cal K$.  A subset of these transformations keeps the form of the $K$-matrix as in \eqref{KforcalK} and does not change the braiding of anyons.  These are transformations of the form
\ie\label{Ktranss}
K  \to \begin{pmatrix}
        1 &&D{\bf L}^\mathrm{T} \\
        0 && 1
    \end{pmatrix} K \begin{pmatrix}
        1 && 0 \\
        {\bf L}D && 1
    \end{pmatrix} \qquad, \qquad {\bf L}\in \bZ^{N\times N}\,.
\fe
They shift $B\to B-{\bf L}-{\bf L}^\mathrm{T}$ and map $p_{ij}$ such that their invariant information, $p_{ij}\bmod \gcd(n_i,n_j)$ for $i\ne j$ and $p_i\bmod 2n_i$, does not change.  (See the discussion after \eqref{braidele}.)

Later, we will need (see \eqref{BMDdef})
\ie\label{DinM}
&\left(BD\right)_{~i}^{j}={p_{ij}n_i\over \gcd(n_i,n_j) }\in \bZ\,,\\
&\left(DB\right)_{i}^{~j}=
        {n_ip_{ij}\over \gcd(n_i,n_j)}\in \bZ\,,
\fe
and
\begin{equation}
    K^{-1}=\begin{pmatrix}
        0 & D^{-1}\\
        D^{-1} & B
    \end{pmatrix}\,.
\end{equation}

The set of anyons
\ie\label{KAv}
{\cal K_A} =\left\{\begin{pmatrix} 0 \\ \omega_c \end{pmatrix} \right\}
\fe
braid as $\omega_c^\mathrm{T} B\omega'_{ c}$.  Given the identifications \eqref{omegaident}, these anyons realize ${\cA}$. Therefore, we denote them as ${\cal K_A}\subset{\cal K}$.

The commutant of ${\cal K}_\cA$ (i.e., the elements in $\cal K$ that braid trivially with $\cal K_A$) is
\begin{equation}\label{vecAb}
   {\cal K_{\bar A}} =\left\{ \begin{pmatrix}
        DB       \bar\omega_c\\
        -\bar\omega_c
    \end{pmatrix} \right\}\,.
\end{equation}
The braiding of its elements is $-        \bar\omega_c^\mathrm{T} B \bar\omega'_{ c}$. Given the identifications \eqref{omegaident}, these anyons realize $\bar {\cA}$.
Therefore, we denote them as ${\cal K}_{\bar \cA}\subset {\cal K}$ and hence the bar over $\omega_c$.

The two sets of anyons,  ${\cal K}_{ \cA}$ of \eqref{KAv} and ${\cal K}_{\bar \cA}$ of \eqref{vecAb} might not be distinct.  In particular, the elements ${\cal K}_{\cA\cap\bar \cA}\subset {\cal K}_{\bar \cA}$ with $\bar\omega_c$ such that $B\bar\omega_c\in \bZ^N$ are also in ${\cal K}_{\cA}$.  Denote the set of anyons generated by ${\cal K}_{\cA}$ and ${\cal K}_{\bar \cA}$ as ${\cal K}_{\cA\cup \bar{\cA}}\subseteq {\cal K}$.  The anyons in ${\cal K}_{\cA\cap\bar \cA}$ braid trivially with all the elements in  ${\cal K}_{\cA\cup \bar{\cA}}$. Therefore, if ${\cal K}_{\cA\cap \bar\cA}$ contains nontrivial elements,  ${\cal K}_{\cA\cup \bar{\cA}}$ is not modular. And the elements in $\cal K$ that are not in ${\cal K}_{\cA\cup \bar{\cA}}$ braid nontrivially with ${\cal K}_{\cA\cup \bar{\cA}}\subset {\cal K}$.   $\cal K$ is a minimal modular extension of ${\cal K}_{\cA\cup \bar{\cA}}$.

In terms of one-form symmetry groups, the symmetry of $\cal K_A$ is ${\cal G}_{\cal A}^{(1)}$, and the symmetry of $\cal K_{\bar A}$ is ${\cal G}_{\cal{\bar A}}^{(1)}$ (which is isomorphic to ${\cal G}_{\cal A}^{(1)}$).  But the symmetry of ${\cal K}_{\cA\cup \bar{\cA}}$ is ${\cal G_A}\times \cal G_{\bar A} \over {\cal G_{A\cap \bar A}}$, where the denominator is the symmetry of $\cal K_{A\cap \bar A}$.

\subsection{Focusing on a particular anyon $q \in {\cal A}\subset {\cal K}$}\label{focusqA}

Here we discuss another presentation of $\cal K$ using a different K-matrix.  As we will see, this presentation is convenient when, as in Section \ref{focusq}, we want to focus on a particular anyon $q\in {\cal A}\subset {\cal K}$.  It will also be convenient for our discussion of the transition theory below.

In addition to the field redefinition \eqref{Ktranss}, we can also describe $\cal K$ by other K-matrices.  In particular, consider making a trivial modification of the symmetry group ${\cal G}_{\cal A}^{(1)}$ by multiplying it with the trivial group $\bZ_1^{(1)}$.  This amounts to changing in the discussion above, $N\to N+1$ with $n_{N+1}=1$.  Then, $p_{i, N+1}={m}_i$ are  arbitrary integers, and $p_{N+1}=p_{N+1,N+1}=P$ is an arbitrary even integer.  Clearly, these modifications do not affect $\cal A$ or the TQFT $\cal K$.  However, the K-matrix \eqref{KforcalK} is now $(2N+2)\times (2N+2)$-dimensional of the form
\ie\label{KforcalKee}
     \begin{pmatrix}
    -DBD && D&&- D{m} && 0\\
    D && 0&& 0&&0\\
       -{m}^\mathrm{T}D && 0&& -P && 1\\
        0&&0 && 1&& 0
    \end{pmatrix} \qquad, \qquad {m}\in \bZ^N\qquad, \qquad P\in 2\bZ\,,
\fe
where the order for the rows and the columns is $b^i$, $c^i$, $b^{N+1}=\hat b$, $c^{N+1}=\hat c$.  And of course, we can add more nonzero entries in this $K$-matrix without affecting the TQFT $\cal K$.  In particular, below, we will use
\ie\label{KforcalKeer}
     \hat K= \begin{pmatrix}
    -DBD && D&&- D{m} && 0\\
    D && 0&&-q&&0\\
       -{m}^\mathrm{T}D && -q^\mathrm{T}&& -P && 1\\
        0&&0 && 1&& 0
    \end{pmatrix} \qquad, \qquad {m},q\in \bZ^N\qquad, \qquad P\in 2\bZ\,.
\fe

The redefinition
\ie\label{hatcred}
\hat c\to \hat c+{m}^\mathrm{T}D b +q^\mathrm{T}c+{P\over 2}\hat b
\fe
shifts the parameters to ${m}=q=P=0$, making the theory based on $\hat K$ equivalent to the original one based on $K$ and a decoupled trivial theory.  (Comparing with the redefinitions \eqref{Ktranss}, the shift by ${m}^\mathrm{T}Db +{P\over 2} \hat b$ extends $\bf L$ in \eqref{Ktranss} to a $(N+1)\times (N+1)$-dimensional matrix.  And the shift by $q^\mathrm{T} c$ redefines the generators of the one-form symmetry.)

The reader might be surprised that we consider replacing the $2N\times 2N$-dimensional matrix $K$ of \eqref{KforcalK} with the more complicated $(2N+2)\times (2N+2)$-dimensional matrix $\hat K$  of \eqref{KforcalKeer}.  After all, they both lead to the same TQFT $\cal K$.  The reason for that is that this presentation of $\cal K$ is convenient when, as in Section \ref{focusq}, we want to focus on a particular anyon $q\in {\cal A}\subset {\cal K}$.

Consider the anyon $e^{i\int \hat c}$ corresponding to the charge vector $\begin{pmatrix}0\\0\\0\\1\end{pmatrix}$ with the K-matrix $\hat K$ of \eqref{KforcalKeer}.  The redefinition \eqref{hatcred} maps
\ie\label{maphatc}
&\hat K= \begin{pmatrix}
    -DBD && D&&- D{m} && 0\\
    D && 0&&-q&&0\\
       -{m}^\mathrm{T}D && -q^\mathrm{T}&& -P && 1\\
        0&&0 && 1&& 0
\end{pmatrix}\quad \longrightarrow \quad \begin{pmatrix}
    -DBD && D&&0 && 0\\
    D && 0&&0&&0\\
       0 && 0&& 0 && 1\\
        0&&0 && 1&& 0
    \end{pmatrix} \\
&\begin{pmatrix}0\\0\\0\\1\end{pmatrix}\quad \longrightarrow\quad\begin{pmatrix}Dm \\q\\{P\over 2}\\1\end{pmatrix}
\sim\begin{pmatrix}Dm \\q\\{P\over 2}\\1\end{pmatrix} +\begin{pmatrix}
    -DBD && D&&0 && 0\\
    D && 0&&0&&0\\
       0 && 0&& 0 && 1\\
        0&&0 && 1&& 0\end{pmatrix} \begin{pmatrix}0\\-m\\-1\\-{P\over 2}\end{pmatrix}=\begin{pmatrix}0 \\q\\0\\0\end{pmatrix}\,.
\fe
We see that in terms of $\hat K$, the anyon $q\in {\cal A}\subset {\cal K}$ has a particularly simple presentation as $e^{i\int \hat c}$.

Another way to understand it is the following.  Just as we peeled off one Abelian anyon in \cite{Cheng:2026qax}, and we peeled off all the Abelian anyons in Section \ref{Abelian anyons}, we can start with the TQFT $\cal K$, present it using the K-matrix $K$ of \eqref{KforcalK}, and peel off the anyon $q$ out of it. To relate it to the discussion of $\hat K$ of \eqref{KforcalKeer}, write\footnote{Our notation is similar to \cite{Cheng:2026qax}. $b$, $c$, $\hat b$, etc. are ordinary $\U$ gauge fields and $\check b$, $\check c$, $\check{\hat b}$ are twisted gauge fields. (See below.)  We apologize if this looks ugly. }
\ie\label{primefi}
&b={\check b}+n D^{-1} q \check {\hat b}\\
&c={\check c}+n(Bq+m)\check {\hat b}\\
&\hat b=n\check {\hat b}\,,
\fe
to express the Lagrangian as
\ie\label{LagwithP}
&{\cal L}={1\over 4\pi}\begin{pmatrix}
    {\check b}&{\check c}\end{pmatrix}K \begin{pmatrix}
        d{\check b}\\d{\check c}\end{pmatrix}+{\cal L}_{{\cal T}_{n,p}}\\
&{\cal L}_{{\cal T}_{n,p}}=-{pn\over 4\pi} \check {\hat b}d\check {\hat b}+{n\over 2\pi}\check {\hat b}d\hat c\\ &p=nP+2nm^\mathrm{T}q+nq^\mathrm{T}Bq\,.
    \fe
Following \eqref{defbe} and \eqref{uofqd}, the coefficients in \eqref{primefi} are quantized.  However, they do not correspond to a $\mathrm{GL}(2N+2,\bZ)$ transformation.  This means that ${\check b}$, ${\check c}$, and $\check {\hat b}$ are twisted gauge fields with fluxes
\ie\label{primefif}
& \int {d {\check b}\over 2\pi }+n D^{-1} q \int {d\check {\hat b}\over 2\pi }\in \bZ\\
& \int{d {\check c}\over 2\pi }+nBq \int  {d\check {\hat b}\over 2\pi }\in \bZ\\
& \int {d \check {\hat b}\over 2\pi }\in {1\over n}\bZ\,.
\fe
These twisted fields represent the $\bZ^{(1)}_n$ gauging in
\ie\label{quotientK}
&{{\cal K}\boxtimes {\cal T}_{n,p}\over \bZ_n}\cong {\cal K}
\,.
\fe
Here, the left-hand side is described by the Lagrangian \eqref{LagwithP} with its twisted gauge fields, and the right-hand side represents the original Lagrangian with $K$.  See \cite{Cheng:2026qax} for more details.  In particular, the quotient in \eqref{quotientK} is by the $\bZ_n^{(1)}$ one-form symmetry generated by
\ie\label{anyoninquoK}
\exp\left(i\oint \left(q^\mathrm{T} \check c + p \check {\hat b} -\hat c\right)\right)\,.
\fe

All this means that the choice of $q$ is captured by $\hat K$ rather than by the Wilson line.  Also, $\hat K$ depends on additional parameters $ m$ and $P$ that do not affect the TQFT.  These facts will be important below.

\subsection{Back to $\cal T$}

Consider the TQFT
\ie\label{TKgauget}
{{\cal T}\boxtimes {\cal K}\over {\cal G_A}}\,,
\fe
where $\cal G_A$ in the denominator is the anomaly-free $\cal G_A$, which is the diagonal subgroup of $\cal G_A$ acting on $\cal T$ and $\cal G_{\bar A}$ acting on $\cal K$.\footnote{Here, the quotient means that we gauge the one-form symmetry $\cal G_A$ in the denominator.  See \cite{Cheng:2026qax} for a review of gauging a one-form symmetry.  That review includes a summary of the subject's history (with appropriate references) and outlines how to apply the gauging procedure.  In particular, it includes three steps: removing non-invariant lines, identifying lines that differ by fusing with a gauge line, and treating the fixed point of the identification more carefully.}  To determine this gauged TQFT, we repeat the discussion for $N=1$ in \cite{Cheng:2026qax}.  We denote the elements in the numerator by $(s,\omega_b,\omega_c)$.  Then, rule two of the gauging (identification of lines that differ by fusing with a gauge line) allows us to set $\omega_c=0$. In this case, the identification has no fixed point, and therefore, there is no need to resolve it. And finally, rule one tells us to keep only the invariant lines and hence $(\omega_b)_i=n_s^i\bmod n_i$.  
The conclusion is that \eqref{TKgauget} is isomorphic to $\cal T$
\ie\label{TKgaugetc}
{\cal T} \cong {{\cal T}\boxtimes {\cal K}\over {\cal G_A}}\,.
\fe

This gauging gives another reason to focus on the particular ${\rm GL}(2N,\bZ)$ transformation in \eqref{Ktranss}.  It corresponds to the map
\ie\label{cshiftedbyg}
\begin{pmatrix}
        b \\
        c
    \end{pmatrix} \to \begin{pmatrix}
        1 && 0 \\
        {\bf L}D && 1
\end{pmatrix}\begin{pmatrix}
        b \\
        c
    \end{pmatrix}\,.
\fe
Under this transformation, $b$ does not transform, and $c$ is shifted by an amount proportional to $Db$.  These facts are such that this transformation does not interfere with the quotient in \eqref{TKgaugetc}.

In \cite{Cheng:2026qax} (and also around \eqref{primefi}), we presented the gauging for $N=1$ using twisted gauge fields.  It is straightforward to repeat the same analysis here.  The anyons generating the gauging in \eqref{TKgauget} are products of Abelian anyons in $\cal T$ and anyons in ${\cal K}_{\bar {\cal A}}\subset{ \cal K}$ of \eqref{vecAb}.  The factor from $\cal K_{\bar A}$ twists the $N$ gauge fields $b$.\footnote{In more detail, the $N$ gauge fields $b^i$ are twisted to ${\check b}^i$ with fractional fluxes
\ie
\int {d {\check b}^i\over 2\pi}\in {1\over n_i}\bZ\,.
\fe
The gauge fields ${\frak b}^i=n_i {\check b}^i$ are then ordinary $\U$ gauge fields \cite{Cheng:2026qax}. \label{frakbd}}  But the $N$ gauge fields $c$ remain smooth.  Therefore, the presentation of $\cal T$ as the right-hand side of \eqref{TKgaugetc} allows us to describe the one-form symmetry of $\cal T$ in terms of $N$ ordinary $\U$ gauge fields $c$.

This generalizes the duality in \cite{Cordova:2017vab,Hsin:2018vcg}
\ie\label{peelinganyonfromT}
 {\cal T} \cong {{\cal T}\boxtimes {\cal T}_{n,p}\over \bZ_n}\,,
\fe
which exposes one Abelian anyon generating  $\bZ_n^{(1)}$ with anomaly $p$, to any Abelian group with any anomaly.

Finally, it is straightforward to repeat this discussion using the presentation of $\cal K$ in terms of the K-matrix $\hat K$ of \eqref{KforcalKeer}.

\subsection{Another perspective on expressing $\cal T$ in terms of $\cal K$}\label{Anotherdescription}

Here we present another description of \eqref{TKgauget}, generalizing results in \cite{Cheng:2022nds, Cheng:2026qax}.

We start with the TQFT $\cal T$, with its one-form symmetry $\cal G_A$ with background gauge fields ${\frak B}^i$ (see \eqref{anomalytheory}).  We enrich the theory by $\U^N$ background gauge fields ${\bf B}^i$ by writing
\ie
{\frak B}^i=\left[{d{\bf B}^i\over 2\pi}\right]_{n_i}\,.
\fe
Here, the left-hand side is a two-cocycle modulo $n_i$, which is determined using the background $d{\bf B}$.  With these ${\bf B}$ background fields, we can also add counterterms\footnote{Note, $\frak B$ are two-cocycles for ${\cal G_A}$.  ${\bf B}$ are classical $\U^N$ gauge fields.  And above and below, $B$ is an $N\times N$ braiding matrix.}
\ie\label{bfBcounter}
{1\over 4\pi} {\bf B}^\mathrm{T} \left({\bf L}+{\bf L}^\mathrm{T}\right)d{\bf B} \qquad , \qquad {\bf L}\in \bZ^{N\times N}\,.
\fe
(This $\bf L$ will soon be related to $\bf L$ in \eqref{Bshift}.)

So far, we have not changed $\cal T$; we have simply coupled it to the background ${\bf B}$.  Next, we make $\bf B$ dynamical and denote it $\frak b$.  Here, it is convenient to write ${\bf B}=\frak b=D {\check b}$ with ${\check b}$ twisted dynamical gauge fields, whose fractional fluxes are correlated with the ${\cal G}_{\cal A}^{(1)}$ twists in $\cal T$ through $\int {\frak B}^i=\left(n_i\int {d {\check b}^i\over 2\pi} \right)\bmod n_i$.  (Note that as in footnote \ref{frakbd}, $\int {d {\check b}^i\over 2\pi} \in {1\over n_i}\bZ$, so this equation is nontrivial.) Using these fields, as opposed to the standard $\U$ gauge fields $D{\check b}$, allows us to remove their direct couplings to the fields in $\cal T$.  The only coupling between them arises from the fact that the fractional fluxes of $ {\check b}$ couple to $\cal T$.  This is similar to the change of variables \eqref{primefi} that removes the off-diagonal terms in the K-matrix at the cost of having gauge fields with fractional fluxes. See \cite{Cheng:2022nds, Cheng:2026qax} for more details.

The result of making these gauge fields dynamical is to change\footnote{Strictly, this result is valid only when $\det  K_0\ne 0$.  Otherwise, we need to add kinetic terms for the dynamical gauge fields.\label{Addkineticcheck}}
\ie
{\cal T} \to {\cal T}_0={{\cal T}\boxtimes {\cal K}_0 \over {\cal G_A}}\,,
\fe
where ${\cal K}_0$ is a $\U^N$ Chern-Simons theory with the K-matrix
\ie\label{checkK}
 K_0=-D\Big(B-\left({\bf L}+{\bf L}^\mathrm{T}\right)\Big)D\,.
\fe

If we study the theory in the numerator, the TQFT ${\cal K}_0$ includes ordinary $\U^N$ gauge fields $b$.  (The quotient makes them twisted gauge fields $ {\check b}$.)  To see that they are coupled through the K-matrix \eqref{checkK}, we used the anomaly expression \eqref{anomalytheory} and the counterterm \eqref{bfBcounter}.

Next, we gauge the quantum $\U^N$ zero-form symmetry by adding ${1\over 2\pi} {{\check b}}^\mathrm{T} Ddc$ to recover
\ie
{\cal T}\cong{{\cal T}\boxtimes {\cal K} \over {\cal G_A}}\,.
\fe
(This conclusion is also valid when $\det  K_0=0$ without a kinetic term for the gauge fields.  Compare with footnote \ref{Addkineticcheck}.)

Importantly, the TQFT $\cal T$ is independent of $\bf L$.  But the TQFTs $ {\cal K}_0$ and ${\cal T}_0= {{\cal T}\boxtimes {\cal K}_0 \over {\cal G_A}}$ do depend on that data.

\section{Proliferation transitions}\label{proliferationtransitions}

\subsection{The transition theory}

In this section, we couple the TQFT $\cal T$ to a single complex scalar field $\Phi$.  We will do that by presenting $\cal T$ as \eqref{TKgaugetc}
\ie\label{TKgaugetcc}
{\cal T} \cong {{\cal T}\boxtimes {\cal K}\over {\cal G_A}}\,,
\fe
and coupling $\Phi$ to the Abelian fields that describe $\cal K$.

Specifically, the scalar field $\Phi$ is coupled to the gauge fields via the covariant derivative\footnote{We use $\mathrm{D}$ for the covariant derivative, and $D$ for the diagonal matrix of integers $D_{ij}=n_i\delta_{ij}$.  We hope that this will not cause confusion.} $\mathrm{D}_f\Phi$, where $f$ is a linear combination with integer coefficients of the Abelian fields in $\cal K$.  Importantly, this linear combination should be such that it is compatible with the one-form gauging by $\cal G_A$ in the denominator of \eqref{TKgaugetcc}.  Therefore, our Lagrangian is of the form
\ie
{\cal L_T} +|\mathrm{D}_f\Phi|^2 -\mu^2|\Phi|^2+\cdots\,,
\fe
where ${\cal L_T}$ is a Chern-Simons Lagrangian for $\cal T$ and the ellipses represent higher order terms.

As a first thought, we could try to present $\cal K$ using the K-matrix \eqref{KforcalK}
\ie\label{KforcalKa}
    K = \begin{pmatrix}
        -DBD && D \\
        D && 0
    \end{pmatrix}\,
\fe
and couple $\Phi$ to the gauge field
\ie\label{fdefinition}
&f= {m}^\mathrm{T}D b+ q^\mathrm{T} c \qquad ,\qquad q,{m}\in \bZ^N\,.
\fe
This coupling means that $\Phi$
creates the anyon $q$.

We note that the coupling to $b$ via $Db$ makes it compatible with the $\cal G_A$ gauging.  Also, different values of the integers $m$ and $q$ that correspond to the same anyon type can lead to different field theories.  Below, we will analyze this dependence in detail.

Even though this construction, which is based on the K-matrix $K$, is natural, it does not lead to the most general coupling.  Instead, we will represent $\cal K$ using a larger K-matrix  \eqref{KforcalKeer}
\ie\label{KforcalKeerA}
     \hat K= \begin{pmatrix}
    -DBD && D&&- D{m} && 0\\
    D && 0&&-q&&0\\
       -{m}^\mathrm{T}D && -q^\mathrm{T}&& -P && 1\\
        0&&0 && 1&& 0
    \end{pmatrix} \qquad, \qquad {m},q\in \bZ^N\qquad, \qquad P\in 2\bZ\,.
\fe
In this description, $\Phi$ couples only to the gauge field $\hat c$ with strength one, i.e., its covariant derivative is $\mathrm{D}_{\hat c}\Phi$.  All the information about the choice of anyon $q$ is included in the matrix $\hat K$.

The K-matrices \eqref{KforcalKa} and \eqref{KforcalKeerA} describe the same TQFT $\cal K$, and therefore they correspond to the same  TQFT $\cal T$.  However, the two non-topological theories based on \eqref{KforcalKa} and \eqref{KforcalKeerA} might not be the same.  As we will show in Appendix \ref{KvshatK}, depending on the parameters, the one based on $\hat K$ is more general.

\subsection{The independent data}

In Sections \ref{Abelian anyons} and \ref{AintermsofU1}, we discussed only the TQFT $\cal T$ and the analysis depended only on the data of its Abelian anyons $\cal A$, namely $p_{ij}\bmod \gcd(n_i,n_j)$ for $i\ne j$ and  $p_i\bmod 2n_i$. (And there was also freedom to change the generators of the one-form symmetry.)

Later, we focused on an Abelian anyon $q\in {\cal A}\subset {\cal K}$.  Even for a given K-matrix, e.g., $K$ or $\hat K$ above, different charge vectors represent the same anyon.

In this section, we discuss the non-topological transition theory.  What are its independent discrete parameters beyond those of $\cal T$ and the choice of anyon $q$?\footnote{Clearly, there are additional continuous parameters like potential terms for $\Phi$.}

As in \eqref{Ktranss}, the matrix $\hat K$ can be changed by GL$(2N+2,\bZ)$ transformations.  We limit ourselves to transformations that do not change the form of the matrix $\hat K$, and we do not transform $\hat c$, which couples to $\Phi$.  Therefore, we study
\ie\label{Ktransse}
&\hat K  \to \hat M^\mathrm{T} \hat K \hat M\\
    &\hat M=\begin{pmatrix}
        1 && 0&&{\bf k}_b&&0 \\
        {\bf L}D && 1&&{\bf k}_c&&0\\
       0&& 0&&1&&0\\
        0&&0&&0&&1
    \end{pmatrix} \qquad, \qquad {\bf L}\in \bZ^{N\times N}\qquad,\qquad {\bf k}_b,{\bf k}_c\in \bZ^N\,,
\fe
i.e., extending \eqref{cshiftedbyg} to
\ie
&b \to b+ {\bf k}_b\hat b\\
&c\to c+ {\bf L}Db+{\bf k}_c\hat b\,.
\fe
This field redefinition changes the entries of the matrix $\hat K$ as follows:
\ie\label{parameterchange}
&B\to B-{\bf L}-{\bf L}^\mathrm{T}\\
&q\to q-D{\bf k}_b\\
&{m}\to {m}-{\bf k}_c+BD{\bf k}_b  +{\bf L}^\mathrm{T}(q-D{\bf k}_b)\\
&P\to P+{\bf k}_b^\mathrm{T}DBD{\bf k}_b -2({\bf k}_b^\mathrm{T} D {\bf k}_c-{\bf k}_b^\mathrm{T}D{m}-{\bf k}_c^\mathrm{T}q)\,,
\fe
where we used the fact that $D$ is diagonal.  As a check, the shift of $P$ is even.  (Compare with footnote \ref{shiftqT}.)

Let us review the various parameters.   We started with ${\cal A}\subseteq {\cal T}$.  It is characterized by $n_i$ and the braiding data $B$, which includes $p_{ij}$.  We also introduced the Chern-Simons couplings with coefficients ${m}$, $q$, and $P$, which do not affect $\cal K$ (and therefore do not affect $\cal T$).  We should mod out this data by the parameters labeling the field redefinitions $\bf L$, ${\bf k}_b$, and ${\bf k}_c$. Depending on how we organize this modding-out, we obtain different parameterizations of the independent data.

As we discussed around \eqref{Ktranss}, we can use $\bf L$ to mod out the redundancy in $p_{ij}$.  So the data in $B$ is the same as in the original TQFT. This is the result of the first equation in \eqref{parameterchange}.  The second equation in \eqref{parameterchange} shows that we can use ${\bf k}_b$ to have $q$ label only the anyon type, i.e., $q\sim q- D {\bf k}_b$.  And the third equation in \eqref{parameterchange} allows us to use ${\bf k}_c$ to set ${m}=0$.

At this stage, the theory is labeled by the following information.
\begin{itemize}
    \item The data of the TQFT, including $p_{ij}\bmod \gcd(n_i,n_j)$ for $i\ne j$ and $p_i=p_{ii}\bmod 2n_i$.
    \item The anyon type of $q$, which is $q \sim q - D{\bf k}_b$.
    \item An arbitrary even integer $P$ (after the lifts of $B$ and $q$ above are fixed).
\end{itemize}

As a check of this picture, consider the spin of the anyon $q$.  Using $\hat K$, we find
\ie\label{hofq}
&h(q)={1\over 2}
\begin{pmatrix}0&0&0&1\end{pmatrix}\hat K^{-1}\begin{pmatrix}0\\0\\0\\1\end{pmatrix}={P\over 2}+{m}^\mathrm{T} q+{1\over 2}q^\mathrm{T}Bq={p\over 2n}\\
&p=nP+2nm^\mathrm{T}q+q^\mathrm{T}u\,.
\fe
(Compare with \eqref{LagwithP}.) At the level of the TQFT $\cal T$, only $h(q)\bmod 1= {1\over 2}q^\mathrm{T}Bq \bmod 1$, and therefore only $p\bmod 2n$ are meaningful.  This result agrees with our discussion of the TQFT in Section \ref{Abeliannotation}.   As we discussed in \cite{Cheng:2026qax}, in the non-topological transition theory, $h(q)$ and $p$, and not only $h(q)\bmod 1$ and $p\bmod 2n$ are meaningful.  It is straightforward to check that $h(q)$ and $p$ of \eqref{hofq} are indeed invariant under \eqref{parameterchange}.

We conclude that the theory depends on the data of $\cal T$, the choice of anyon type $q$ (i.e., the vector $q$ up to identifications), and a single additional integer parameter.
This is consistent with the discussion in \cite{Cheng:2026qax} and extends it to a broader context with additional fields.

As we said, without loss of generality, we can focus on $m=0$.  Below, we will simplify the discussion by setting $m=0$.
\ie\label{KforcalKeerAmz}
     \hat K= \begin{pmatrix}
    -DBD && D&& 0&& 0\\
    D && 0&&-q&&0\\
       0 && -q^\mathrm{T}&& -P && 1\\
        0&&0 && 1&& 0
    \end{pmatrix} \qquad, \qquad q\in \bZ^N\qquad, \qquad P\in 2\bZ\,.
\fe
Then, the available reparameterizations in \eqref{parameterchange} are only those that preserve $m=0$.

\subsection{A dual description}
\label{dualdescription}

Here, we use the fact that the scalar field $\Phi$ couples to a single $\U$ gauge field $\hat c$ with charge one to find its dual description.

Following standard particle/vortex duality, the dual theory includes the TQFT
\ie\label{Tprime}
{\cal T}'\cong{{\cal T}\boxtimes {\cal K}'\over {\cal G_A}}\,,
\fe
where $\cal K'$ is the Abelian TQFT described by the $K$-matrix
\ie\label{KforcalKd}
    K' =  \begin{pmatrix}
    -DBD && D&&0 \\
    D && 0&&-q\\
       0 && -q^\mathrm{T}&& -P\\
    \end{pmatrix} \,.
\fe
(Recall that we set $m=0$.)  This is the same as $\hat K$, except that the last row and the last column corresponding to the gauge field $\hat c$ are absent.

This TQFT is coupled to the dual scalar $\tilde\Phi$ only through the $\U$ gauge field $\hat b$ through the covariant derivative $\mathrm{D}_{\hat b}\tilde \Phi$.

Clearly, this dual theory can be modified by field redefinitions that do not change $\hat b$, which couples to $\tilde \Phi$.  In particular, we can change the parameters as in \eqref{parameterchange}.

There is an important subtlety that arises in this dual theory.  The original K-matrix $K$ or $\hat K$ was not singular, $\det K=-\det \hat K=(-1)^N(\det D)^2\ne 0$.  Therefore, we could neglect the gauge kinetic terms.  This is not the case with the dual theory.  Here, the determinant is 
\ie\label{Kprimeh}
\det K'=2(-1)^{N+1}(\det D)^2 h(q)
\fe
with $h(q)$ given by \eqref{hofq}. This determinant can vanish. In that case, the duality is valid only if we include gauge kinetic terms.  Below, when we discuss the phases of the theory, the relation \eqref{Kprimeh} and its zero will be significant.

Again, this dual picture is consistent with the discussion in \cite{Cheng:2026qax} and generalizes it to the broader context in this paper.

\section{The global symmetry of the transition theory}\label{globalU1}

\subsection{The one-form symmetry}\label{oneformoftt}

As in \cite{Cheng:2026qax}, the one-form symmetry of $\cal T$, ${\cal G}_{\cal A}^{(1)}$, which is generated by $\cal A\subseteq T$, is explicitly broken to ${\cal G}_{{\cal A}_q^\perp}^{(1)}$, which is generated by the Abelian anyons that braid trivially with $q$, ${\cal A}_q^\perp\subseteq{\cal A}$.

In terms of the field theory, $\cA_q^\perp$ can be represented by the symmetry operators
\ie
&e^{ir^\mathrm{T}\oint (c- Bq\hat b)}\\
&r\in \cA_q^\perp={\{r\in \bZ^N| r^\mathrm{T}Bq\in \bZ\}\over (r \sim r-D{\bf k})}
\qquad, \qquad {\bf k} \in \bZ^N\,.
\fe
See \eqref{Aperpd}.\footnote{The equations of motion
\eqref{solvingflu} and \eqref{bceomn} show that $dc=Bqd\hat b$, and hence this operator is indeed topological.}
And as in \eqref{ZsubL}, in general, $q$ is not included in it.  Only
\ie\label{ZLinoneform}
\bZ_L^{(1)}\subseteq {\cal G}^{(1)}_{{\cal A}_q^\perp}\,
\fe
generated by $q^{n\over L}$ is included.

\subsection{The (on-shell) current of the zero-form symmetry $\U_{\cal B}$}

Let us turn to the zero-form symmetry. Unlike the TQFT $\cal T$, the transition theory has a global zero-form symmetry, $\U_{\cal B}$, which acts on $\Phi$ by multiplying it by a phase.
Its current $J_{\cal B}$ is proportional to\footnote{We follow our conventions in \cite{Cheng:2026qax}.  A scalar field $\chi$ with $\U_C$ charge $1$ couples to the gauge field $C$ using the covariant derivative  $\mathrm{D}_C\chi=(d -i C)\chi$.  Then, $\chi({\cal P}) \exp\left({i\int_{\cal P}^{{\cal P}'} C}\right)\bar\chi({\cal P}')$ is gauge invariant.  In these conventions, the $\U_C$ current is $J_C=-{\delta {\cal L}\over \delta C}$.  Therefore, a kinetic term like $|\mathrm{D}_C\chi|^2$ contributes to $J_C$
\ie
\mathscr{J}(\chi)=\star i\chi \overset\leftrightarrow{\mathrm{D}} \bar\chi=\star i(\chi \mathrm{D}\bar\chi-\bar\chi \mathrm{D}\chi)\,,
\fe
where $\mathrm{D}$ is the appropriate covariant derivative in each term.  Here, $\star$ is the Hodge dual.  Note that we define the current as a differential form.  People unfamiliar with this notation can simply consider the Hodge dual of our expressions.  In our conventions, a Chern-Simons term, ${k\over 2\pi} C dX$ contributes $J_C=- {k\over 2\pi} dX $.\label{gaugeconv}}
\ie
\mathscr{J}(\Phi)=\star i\Phi \overset\leftrightarrow{\mathrm{D}} \bar\Phi\,.
\fe
Since $\Phi$ is not gauge invariant, the current $J_{\cal B}$ can be a fraction of $\mathscr{J}(\Phi)$.  This fraction should be chosen so that all gauge-invariant operators have quantized charges.\footnote{This situation is analogous to the statement that in QCD, the quarks have baryon charge $1\over 3$, while all the gauge-invariant operators have integer baryon number.}

To find the properly normalized current, we look for a linear combination with integer coefficients of $db\over 2\pi$, $dc\over 2\pi$, $d\hat b\over 2\pi$, $d\hat c\over 2\pi$, and $\mathscr{J}(\Phi)$.  On-shell, we can simplify it using the equations of motion.  The equations of motion of $b$, $c$, $\hat b$ and $\hat c$ are respectively
\ie\label{eomPhi}
&Ddc-DBD db=0\\
&Ddb=qd\hat b \\
&d\hat c=  q^\mathrm{T} dc +P d\hat b \\
&d\hat b =2\pi \mathscr{J}(\Phi)\,.
\fe
(Recall that we set $m=0$.)
Locally, we can solve for all the fluxes in terms of $\mathscr{J}(\Phi)$.  But that might not be compatible with the flux quantization, so we should be more careful.  We can clearly solve for $d\hat c$ and $d\hat b$.  Their quantization does not constrain the other fluxes.  The quotient that we need to perform to find $\cal T$ means that the fluxes $Ddb$ have standard quantization.  Therefore, we solve
\ie\label{solvingflu}
&Ddb=2\pi q\mathscr{J}(\Phi) \\
&d\hat c =2\pi P \mathscr{J}(\Phi) + q^\mathrm{T} dc  \\
&d\hat b =2\pi \mathscr{J}(\Phi)\,.
\fe

We are left with one equation of motion
\ie\label{bceomn}
E={dc\over 2\pi}- Bq \mathscr{J}(\Phi)=0\,.
\fe
It states that the flux quantization of $c$ constrains the eigenvalues of the charge $\int \mathscr{J}(\Phi)$.

Now, let us return to the current.  As we said, the current $J_{\cal B}$ is a linear combination with integer coefficients of $db\over 2\pi$, $dc\over 2\pi$, $d\hat b\over 2\pi$, $d\hat c\over 2\pi$, and $\mathscr{J}(\Phi)$.  Using \eqref{solvingflu}, on-shell, it is a linear combination with integer coefficients of $d c\over 2\pi$ and $\mathscr{J}(\Phi)$
\ie\label{JBcPhi}
J_{\cal B}={1\over 2\pi}t ^\mathrm{T}dc+z\mathscr{J}(\Phi) \qquad, \qquad t\in \bZ^N\qquad, \qquad z\in \bZ\,.
\fe
Using \eqref{bceomn}, on-shell, it is
\ie\label{JBshiftedbyE}
&J_{\cal B}={1\over 2\pi}t^\mathrm{T} dc+z\mathscr{J}(\Phi)-t^\mathrm{T} E={1\over n}\left(t^\mathrm{T}u+z n\right)\mathscr{J}(\Phi)\\
&u=n Bq\,.
\fe
Using $\gcd(n,u^1,u^2,\cdots,u^N)=\ell$ \eqref{elldefu}, the minimal nonzero coefficient of the current arises from a solution of
\ie\label{QcQB}
t^\mathrm{T}u+z n=\ell\,.
\fe
This means that we can choose $t$ to be a particular $\tau$, as in  \eqref{xuell}, with an appropriate integer $z$.  Then, on-shell, the current is
\ie\label{JBc}
J_{\cal B}={1\over 2\pi}\tau^{\mathrm T} dc +z \mathscr{J}(\Phi)={\ell\over n} \mathscr{J}(\Phi) \,.
\fe

We noted above that $\tau$ represents a quotient \eqref{Aquotient}.  Other values of $\tau$ can also be used, provided they satisfy \eqref{tauplusr}. Starting with a particular $\tau$, we can shift $t=\tau \to \tau +r$ with any $r\in {\cal A}_q^\perp$.  Indeed, if we accompany this shift with $z\to z-r^\mathrm{T}Bq$,
equation \eqref{QcQB} is preserved, and the current \eqref{JBc} is shifted by a multiple of the equation of motion $E$ of \eqref{bceomn}.  Consequently, this change in $\tau$ leads to different off-shell expressions for the current \eqref{JBc}, but on-shell, they are equivalent and given by ${\ell\over n} \mathscr{J}(\Phi)$.  This will be discussed further in Section \ref{BackgroundB}.

The first form of the current in \eqref{JBc} makes it clear that its charges are manifestly quantized even off-shell.   The second form of the current in \eqref{JBc} means that we can ignore all the magnetic flux contributions to the current, and instead, think of the scalar field $\Phi$ as having fractional $\U_{\cal B}$ charge given by $\ell\over n$.  Importantly, $\Phi$ is not gauge invariant,  and therefore, its global symmetry charge is ill-defined.  (Indeed, we can change it using transformations, as above.)  However, this statement is meaningful when we consider gauge-invariant operators.  As we will see below, they all have integer $\U_{\cal B}$ charges.

\subsection{Coupling the system to a background $\U_{\cal B}$ gauge field}\label{BackgroundB}

Here we present another perspective on the $\U_{\cal B}$ symmetry.

We started with the TQFT $\cal T$.  This theory has no zero-form continuous symmetry, but it can be enriched by a global $\U$ zero-form symmetry   \cite{barkeshli_symmetry_2019}.  This means that the background gauge fields $\frak B$ of the one-form symmetry ${\cal G}_{\cal A}^{(1)}$ (see \eqref{anomalytheory}) are expressed in terms of an ordinary $\U_{\cal B}$ background gauge field $\cal B$.  Specifically, we couple $\cal T$ to $\cal B$ via the terms  
\ie\label{BtoT}
&-{1\over 2\pi} {\cal B} t^\mathrm{T} dc - {R\over 2\pi} {\cal B}d{\cal B}\\
&t\sim t +D{\bf k}\in \bZ^N\qquad, \qquad  {\bf k}\in \bZ^N\qquad,\qquad R\in \bZ \,.
\fe
Here, the last term is a counterterm in the background field, which can be thought of as representing an SPT phase.

One might try to interpret \eqref{BtoT} as saying that the gauge field ${\cal B}$ couples to the current ${1\over 2\pi}  t^\mathrm{T} dc +{R\over \pi}d{\cal B}$. (See our conventions in footnote \ref{gaugeconv}.) However, using the equations of motion, this expression is a pure c-number and is a trivial operator; this $\U_{\cal B}$ current vanishes on-shell.

Instead, the coupling \eqref{BtoT} means that $t\in {\cal A}$ is a vison for $\U_{\cal B}$; i.e., every anyon $s\in \cal T$ carries $\U_{\cal B}$ charge that is determined by braiding it with $\exp\left(i t^{\mathrm T}\int c \right)$.  This charge is given by $\sum_i {1\over n_i} t_i n^i_s\bmod 1 $.

One might ask whether we can have a more general coupling of $\cal B$ by using the fields $b$, $\hat b$, and $\hat c$ in \eqref{KforcalKeerAmz}, i.e., write
\ie\label{generalBbc}
-{1\over 2\pi} {\cal B} \begin{pmatrix}t_b^\mathrm{T}& t_c^\mathrm{T} & t_{\hat b}^\mathrm{T}& t_{\hat c}^\mathrm{T}\end{pmatrix} \begin{pmatrix}db\\ dc\\ d\hat b\\ d\hat c \end{pmatrix} -{R\over 2\pi}{\cal B}d{\cal B}\,.
\fe
To answer this question, we first note that in order to be compatible with the one-form gauging in \eqref{TKgaugetcc}, we should have $t_b=D {\bf l}$ with ${\bf l}\in\bZ^N$.  Then, we can simplify the couplings to $\cal B$ by shifting the dynamical fields by integer multiples of $\cal B$.\footnote{If we shifted the dynamical fields by $\cal B$ with arbitrary real coefficients, this would have been the same as simplifying the coupling by using the equations of motion.  The restriction to integer coefficients makes this procedure more restrictive than simply using the equations of motion. Below, we will discuss it further.\label{eomvs}}
Specifically, shifting
\ie\label{shiftbyt}
&b\to b-{\cal B} {\bf k}\\
&c\to c+{\cal B}\left({\bf l} -BD{\bf k}\right)\\
&\hat b\to \hat b+{\cal B} t_{\hat c}\\
&\hat c \to \hat c +{\cal B}\left(t_{\hat b}+q^\mathrm{T} {\bf l} -q^\mathrm{T}BD{\bf k}+ Pt_{\hat c}\right)\\
&{\bf k}\in \bZ^N
\fe
maps  $t_b,\ t_{\hat b},\ t_{\hat c} \to 0$, and $t_c\to t_c+qt_{\hat c} + D{\bf k}$.  Denoting $t=t_c+qt_{\hat c}$, we see that, up to field redefinitions and higher order terms, the most general coupling is \eqref{BtoT}.  (This also explains the identification in \eqref{BtoT}.)

Next, we couple our system to $\Phi$ and turn on the background gauge field $\cal B$.  In addition to the coupling \eqref{BtoT}, now, we also have another coupling of $\cal B$ through the covariant derivative of $\Phi$, given by
\ie\label{hatcBco}
\mathrm{D}_{\hat c+z{\cal B}}\Phi\qquad , \qquad z\in \bZ\,.
\fe
Since the identification  $t\sim t+D{\bf k}$ in \eqref{BtoT} involved shifting $\hat c$ by an integer multiple of $\cal B$ according to \eqref{shiftbyt}, now, it becomes
\ie\label{geniden}
(t,z)\sim (t+D{\bf k},z-{\bf k}^\mathrm{T}DBq)\,.
\fe
Note that the shift of $z$ is by an integer.

So far, we determined the most general coupling of $\cal B$.  As above, the minimal charge for $\U_{\cal B}$ is obtained when we impose \eqref{QcQB}
\ie\label{QcQBg}
t^\mathrm{T}u+z n=\ell\,.
\fe
Note that it is consistent with the identification in \eqref{geniden}.

We conclude that, up to field redefinitions and higher order terms,  $\cal B$ couples as
\ie\label{BtoTPhi}
&-{\cal B}\left({1\over 2\pi}t^\mathrm{T} dc+z\mathscr{J}(\Phi)\right)\\
&t\in \bZ^N\qquad,\qquad z\in \bZ \\
&t^\mathrm{T}u+z n=\ell\\
&(t,z)\sim (t+D{\bf k},z-{\bf k}^\mathrm{T}DBq)\,.
\fe

This seems similar to the discussion leading to \eqref{JBc}.  However, there is an important difference between them.  In \eqref{JBshiftedbyE}, we used the equations of motion \eqref{bceomn} and shifted the current by $t^\mathrm{T}E$, which vanishes on-shell.  This is correct when we discuss the operator $J_{\cal B}$. Instead, now we can shift the dynamical fields only by integer multiples of ${\cal B}$.  This maps \eqref{BtoTPhi}
\ie\label{BtoTPhis}
-{\cal B}\left({1\over 2\pi}t^\mathrm{T} dc+z\mathscr{J}(\Phi)\right) \to -{\cal B}\left({1\over 2\pi}t^\mathrm{T} dc+z\mathscr{J}(\Phi)+{\bf k}^\mathrm{T} D E\right)\,,
\fe
and leads to the identification \eqref{geniden}.  Therefore, the on-shell current is
\ie\label{oneshellJ}
J_{\cal B}={1\over 2\pi}t^\mathrm{T} dc+z\mathscr{J}(\Phi)-x^\mathrm{T} E\,,
\fe
with arbitrary $x\in \bR^N$, but the coupling to the background $\cal B$ is restricted to $x=D{\bf k}$ with ${\bf k}\in \bZ^N$.

Let us count the different parameters.  The on-shell current is unique and is given by \eqref{JBc}.  The coupling to $\cal B$ is labeled by $t\in \bZ^N$ and $z\in \bZ$ subject to the relation and the identifications in  \eqref{BtoTPhi}.  Another way to describe it is as follows.  Above, we saw that the minimal charge constraint \eqref{QcQBg}
can be solved with $t=\tau$ with a particular $\tau$, as in \eqref{xuell}, with an appropriate integer $z$.  Other solutions of this equation are obtained by
\ie\label{tQPhishift}
&t=\tau\to \tau +r\\
&z \to z -r^\mathrm{T}B q\\
&r\in {\cal A}_q^\perp\,.
\fe
Here, the restriction to $r\in {\cal A}_q^\perp$ guarantees that $r^\mathrm{T}Bq\in \bZ$, and therefore, the shift of $z$ is by an integer.  These correspond to shifting the current as in \eqref{oneshellJ} with $x=-r$.    Importantly, these shifts by $r$, which are trivial on-shell, lead to distinct couplings of $\cal B$ because nontrivial $r$ is not of the form $D{\bf k}$.

To summarize, the coupling to $\cal B$ involves a coupling to the on-shell current $J_{\cal B}$, as well as terms labeled by $r\in {\cal A}_q^\perp$.  These terms, which vanish on-shell, represent enrichment using the one-form symmetry of the transition theory ${\cal A}_q^\perp$.  (See the comment after \eqref{JBc}.)

In other words, the presence of $\Phi$ explicitly breaks ${\cal G}_{\cal A}^{(1)}$ to ${\cal G}_{{\cal A}_q^\perp}^{(1)}$. (In the special case of $\ell=n$, there is no such breaking.) Therefore, for $\ell \ne n$, an anyon $\tau\in {\cal A}$, which is not in ${\cal A}_q^\perp$, does not generate a one-form symmetry of the transition theory.  Instead, it is related to a faithful global $\U_{\cal B}$ zero-form symmetry. And in addition, we have enrichment coupling using ${\cal A}_q^\perp$.

\subsection{The charged operators}

Denote the monopole operator of the gauge field $x$ by ${\cal M}_{x}$. Then, consider the operator
\ie\label{COell}
&{\cal O}=\Phi^n\prod_{i=1}^N\left(  {\cal M}_{b^i}^{nq_i\over n_i}{\cal M}_{c^i}^{u^i}\right){\cal M}_{\hat b}^{n}{\cal M}_{\hat c}^{p}\\
&p=2nh(q)=nP+q^\mathrm{T}u\,.
\fe
It is well-defined as all the exponents are integers.  And it is gauge invariant, as can be seen by checking that it satisfies the equations of motion  \eqref{solvingflu} and \eqref{bceomn}.\footnote{Let us express this operator using the twisted gauge fields in \eqref{primefi}
\ie
&{\check b}=b- D^{-1} q \hat b\\
&{\check c}=c-Bq\hat b\\
&\check {\hat b}={1\over n}\hat b\,.
\fe
(Recall that we set $m=0$.)
$\cal O$ has vanishing ${\check b}$ and  ${\check c}$ fluxes, and hence
\ie
{\cal O}=\Phi^n{\cal M}_{\hat b}^{n}{\cal M}_{\hat c}^{p}=\Phi^n{\cal M}_{\check {\hat b}}{\cal M}_{\hat c}^{p}\,.
\fe
(Note that when we write monopole operators like ${\cal M}_{\hat b}$ or ${\cal M}_{\hat c}$, we need to specify which other gauge fields have vanishing fluxes.  In \eqref{COell}, these are ${\cal M}_b$ and ${\cal M}_c$, while here, these are  ${\cal M}_{{\check b}}$ and ${\cal M}_{{\check c}}$.) This agrees with the expression in \cite{Cheng:2026qax}.  In more detail, our gauge fields $\hat b$ and $\hat c$ are denoted there as $\frak b$ and $c$ respectively, and our $\check {\hat b}$ was denoted there as $\check b$.}    It carries charge $\ell$ under $\U_{\cal B}$
\ie\label{QBofO}
Q_{\cal B}({\cal O})=\ell\,.
\fe

If we had studied the theory of $\Phi$ coupled to the TQFT $\cal K$, $\cal O$ would have been the lowest charge, gauge-invariant operator.  Potential lower-charge operators necessarily would have had fractional fluxes, which would have been inconsistent.  However, in our case, $\Phi$ couples to $\cal T$.  The one-form gauging in the expression for $\cal T$ in terms of $\cal K$ \eqref{TKgauget} means that the gauge fields $b^i$ that describe $\cal K$ can have fluxes that are multiples of $1\over n_i$.  This allows for lower charge operators.

Specifically, we can study the operator
\ie\label{COellp}
&{\cal O}'= \Phi^{n\over \ell}\prod_{i=1}^N\left( {\cal M}_{b^i}^{nq_i\over n_i \ell} {\cal M}_{c^i}^{u^i\over \ell}\right){\cal M}_{\hat b}^{n\over \ell}{\cal M}_{\hat c}^{p\over \ell}\\
&\ell=\gcd(n,u^1,u^2,\cdots,u^N)\,.
\fe
The value of $\ell$ guarantees that all the exponents except those of ${\cal M}_{b^i}$ are integers.  And the exponents of ${\cal M}_{b^i}$ are multiples of ${1\over n_i}$.

Clearly, this reasoning about the allowed exponents and the fluxes is identical to the one that led us to the normalization of the current \eqref{JBc}.  Indeed, $\cal O'$ carries the minimal $\U_{\cal B}$ charge,
\ie\label{QBofOp}
Q_{\cal B}({\cal O}')=1\,.
\fe

Let us understand better the difference between $\cal O$ and $\cal O'$.  To define $\cal O$, we remove a point from spacetime, specify these fluxes on a small $S^2$ around it, and add $\Phi^n$.  Such a construction is not possible for $\cal O'$.  Here, the fractional fluxes of $b^i$ mean that the gauge fields of the one-form symmetry in the denominator of \eqref{TKgauget} are nontrivial.  This cannot be the case on $S^2$.  Therefore, $\cal O'$ must act along a line.  This line can be surrounded by a torus or by a higher-genus surface.

We conclude that $\cal O$ is a local operator acting at a point.  $\cal O'$ is not a standard local operator.  It does not act at a point but along a line.  Below, we will provide another interpretation of this fact.

Finally, we note that in the dual description in Section \ref{dualdescription}, the monopole operator \eqref{COell} takes the form
\ie\label{COelld}
{\cal O}=\begin{cases}
    \bar{\tilde \Phi}^p\prod_{i=1}^N\left( {\cal M}_{b^i}^{nq_i\over n_i}{\cal M}_{c^i}^{u^i} \right){\cal M}_{\hat b}^{n}& p\ge 0\\
{\tilde \Phi}^{-p}\prod_{i=1}^N\left( {\cal M}_{b^i}^{nq_i\over n_i}{\cal M}_{c^i}^{u^i} \right){\cal M}_{\hat b}^{n}&p\le 0\,.
\end{cases}
\fe
And similarly for $\cal O'$.

\subsection{Summary and interpretation in terms of $\cal T$}

These conclusions extend the discussion in \cite{Cheng:2026qax}.  The properly quantized zero-form symmetry of the model is $\U_{\cal B}$ with the current \eqref{JBc}.  It acts faithfully on the states in the theory.

For $\ell\ne 1$, this symmetry does not act faithfully on local operators.  Only its quotient $\U_{\cal B}\over \bZ_\ell$ does.  Indeed, the local operator $\cal O$ of \eqref{COell} has charge $\ell $ under $\U_{\cal B}$ and charge one under the quotient $\U_{\cal B}\over \bZ_\ell$.  The operator $\cal O'$ of \eqref{COellp} has charge one under $\U_{\cal B}$.  However, it is not a local operator, because it is supported on closed lines.

One way to understand this conclusion is to trace it back to the original TQFT $\cal T$.  As we said, the field $\Phi$ creates the anyon $q$.  Let us consider some of its powers.  (Compare with \eqref{Zsubell}.)
\begin{itemize}
    \item[$q$] This anyon is created by the field $\Phi$.  It is not gauge-invariant.  To make it gauge-invariant, we can attach a line to it, but then it acts at the end of an open line.
\item[$q^{n\over \ell}$]
The simplest $\U_{\cal B}$-charged, gauge-invariant operator is constructed out of $\Phi^{n\over \ell}$.  It can be dressed by a monopole operator to form the gauge-invariant operator $\cal O'$ of \eqref{COellp}.  This operator creates the anyon $q^{n \over \ell}$.  This anyon braids trivially with $\cal A$, and as we discussed in Section \ref{AonT}, it has a nonzero one-point function on some $\Sigma_g$.  This one-point function corresponds to the operator $\cal O'$.  In more detail, consider $\Sigma_g$ with an attached $\Phi^{n \over \ell}$.  Inserting this $\Sigma_g$ with the anyon in spacetime corresponds to a closed line operator.
\item[$q^n$]
The simplest $\U_{\cal B}$-charged point operator is $\cal O$, which corresponds to the trivial anyon $q^n$.
\end{itemize}

\section{The phases}
\label{sec:phases}

Here, we discuss the phases of the theory and their symmetries.  This discussion generalizes \cite{Cheng:2026qax}, with additional details available there.

The global symmetry of our UV transition theory includes the following.
\begin{itemize}
    \item A ${\cal G}_{{\cal A}_q^\perp}^{(1)}$ (invertible) one-form symmetry.  There can also be non-invertible symmetries, which we will ignore.
    \item A $\U_{\cal B}^{(0)}$ zero-form  symmetry.  There can also be other discrete zero-form symmetries, which we will ignore.
    \item Below, we will deform the theory by turning on the operator $\cal O$.  Then, the continuous $\U_{\cal B}^{(0)}$ symmetry is explicitly broken to  $\bZ_\ell^{(0)}\subset \U_{\cal B}^{(0)} $.
\end{itemize}

\subsection{$\mu^2>0$}

For $\mu^2>0$, $\Phi$ is massive and the low-energy theory is the original TQFT $\cal T$.

Let us discuss the fates of the global symmetries of the UV transition theory.
\begin{itemize}
\item The one-form symmetry ${\cal G}_{{\cal A}_q^\perp}^{(1)}$ of the UV theory is enhanced in the IR to an emergent ${\cal G}_{{\cal A}}^{(1)}$ symmetry of $\cal T$.  (In addition, the IR theory also has noninvertible one-form symmetries associated with $\cal T$.) This one-form symmetry is spontaneously broken \cite{Gaiotto:2014kfa}.
\item  The $\U_{\cal B}$ zero-form symmetry does not act at low energies; it is not realized faithfully in the IR.  Therefore, this symmetry is unbroken.  This also includes its $\bZ_\ell $ subgroup.
\item  Turning on the perturbation $\cal O$ explicitly breaks $\U_{\cal B}$ to $\bZ_\ell$.  This breaking does not affect the IR TQFT $\cal T$.  The one-form symmetry is still enhanced to the emergent ${\cal G}_{{\cal A}}^{(1)}$, which is spontaneously broken, and the UV $\bZ_\ell$ does not act in the IR.
\end{itemize}

\subsection{$\mu^2<0$}

The details of the phase depend on whether $p=2nh(q)$ vanishes.  Following  \eqref{Kprimeh} $\det K'=2 (-1)^{N+1}(\det D)^2 h(q)$, this determines whether $K'$ has a vanishing eigenvalue.

\subsubsection{$p\ne 0$}\label{pnotzeron}

When $h(q)\neq 0$, $\Phi$ Higgses $\U_{\hat c}$.  Semiclassically, both $\Phi$ and $\hat c$ are massive and can be integrated out.  The low-energy theory is a gapped phase with the TQFT
\ie\label{calTprime}
{\cal T'}\cong{{\cal T}\boxtimes {\cal K'}\over {\cal G_A}}\,,
\fe
where $\cal K'$ is the Abelian TQFT based on $K'$ of \eqref{KforcalKd}.

The same conclusion is reached in the dual picture of Section \ref{dualdescription}.  Here, the dual scalar $\tilde \Phi$ is massive, and again, we end up with the TQFT $\cal T'$.

To understand this theory better, we use the change of variables \eqref{primefi}
\ie\label{primefic}
&b={\check b}+n D^{-1} q \check {\hat b}\\
&c={\check c}+n Bq\check {\hat b}\\
&\hat b=n\check {\hat b}\,,
\fe
(recall that we have set $m=0$) to express the Lagrangian of ${\cal K'}$ as
\ie\label{KprimeLag}
&{\cal L}_{\cal K'}={1\over 4\pi}\begin{pmatrix}
    {\check b}&{\check c}\end{pmatrix}K \begin{pmatrix}
        d{\check b}\\d{\check c}\end{pmatrix}-{pn\over 4\pi} \check {\hat b}d\check {\hat b}\,.
    \fe
Alternatively, we find this Lagrangian by substituting $\hat c=0$ in \eqref{LagwithP}.

As in the discussion around \eqref{primefi}, the transformation \eqref{primefic} is not a ${\rm GL}(2N+1,\bZ)$ transformation, and therefore,  these twisted fields represent the $\bZ^{(1)}_n$ gauging in
\ie\label{Kprimed}
 {\cal K}'\cong {{\cal K}\boxtimes \U_{-np}\over \bZ_n}\cong {{\cal K}\boxtimes \U_{-np}\over (q,p)}\,.
\fe
Combining this with \eqref{calTprime}
\ie\label{Tprimed}
{\cal T'}\cong{{\cal T}\boxtimes {\cal K'}\over {\cal G_A}} \cong {{\cal T}\boxtimes {{\cal K} \boxtimes \U_{-np}\over (q,p)}\over {\cal G_A}}\cong  {{\cal T} \boxtimes \U_{-np}\over (q,p)}\,.
\fe

The UV ${\cal G}_{{\cal A}_q^\perp}^{(1)}$ is enhanced to an emergent one-form symmetry group ${\cal G}_{\cA'}^{(1)}$ where $\cA'\subseteq {\cal T}'$ are the Abelian anyons in $\cal T'$.  (Again, there are also noninvertible one-form symmetries associated with $\cal T'$.) It is given by
\ie
    \cA'\cong \frac{\cA\boxtimes \U_{-np}}{(q,p)}\cong\frac{\cA\boxtimes \U_{-np}}{  \bZ_n }\,.
\fe
To justify this expression of $\cA'$, we notice that following the three-step gauging procedure, the gauging by $(q,p)$ in \eqref{Tprimed} does not lead to any ``splitting" of non-Abelian anyons in $\cT$.  Hence, the Abelian anyons in $\cT'$ are just the quotient of those in $\cT\boxtimes \U_{-np}$ by $(q,p)$. This conclusion is also consistent with ${\cal K}'$ of \eqref{Kprimed}.\footnote{In general, given a TQFT $\cal S$ with Abelian anyons ${\cal A}\subseteq{\cal S}$, including an anomaly free $\bZ_m^{(1)}$, we can form the quotient ${\cal S}'\cong {{\cal S}\over \bZ_m}$.  In this case, the Abelian anyons ${\cal A}'\subseteq {\cal S}'$ include those in ${\cal A}\over \bZ_m$, but due to splitting in the gauging process, ${\cal S'}$ can include additional Abelian anyons.   A familiar example is ${\cal S}={\rm SU}(2)_4$ with $m=2$ leading to ${\cal S}'={\mathrm{SU}(2)_4\over \bZ_2}\cong {\rm SU}(3)_1\cong {\cal A}'$.}

Finally, $\cA'$ contains $\cA_q^\perp$, since anyons of the form $(s, 0)$ for $s\in \cA_q^\perp$ survive the quotient. Therefore, the UV ${\cal G}_{\cA_q^\perp}^{(1)}$ is spontaneously broken.

We conclude:
\begin{itemize}
    \item  The entire UV ${\cal G}_{{\cal A}_q^\perp}^{(1)}$ is spontaneously broken.  In particular, $\bZ_L^{(1)}\subseteq {\cal G}_{{\cal A}_q^\perp}^{(1)}$, which is generated by $q^{n\over L}$ is spontaneously broken.  That UV symmetry is enhanced in the IR to an emergent $ {\cal G}_{{\cal A}'}^{(1)}$.
\item In this phase, $\U_{\cal B}$ is  unbroken and does not act at low energies.
\item
Turning on the perturbation $\cal O$ explicitly breaks $ \U_{\cal B}$ to $\bZ_\ell $.  This breaking does not affect the IR TQFT $\cal T'$.  The one-form symmetry is still enhanced to the emergent ${\cal G}_{\cA'}^{(1)}$, and the UV $\bZ_\ell$ does not act in the IR.
\end{itemize}

To summarize, the field theory gives a transition from $\cal T$ to $\cal T'$.  This is a realization of the Abelian hierarchy construction of \cite{HaldaneHierarchy, HalperinHierarchy}.  Related TQFT maps from $\cal T$ to $\cal T'$ were studied by various authors including \cite{Blok:1990mc, Blok:1990an, ReadHierarchy1990, wen1992classification, BondersonSlingerland, LanHierarchy2017, Lan:2019gxv, Hsin:2018vcg, Cheng:2022nds, zhang2024hierarchy, Shi2025}.

\subsubsection{$p= 0$ without adding $\cal O$ to the Lagrangian}

We could attempt to follow the discussion for nonzero $p$ in Section \ref{pnotzeron} to find a theory like $\cal T'$.  However, since $p=0$, $\det K'=0$, and $K'$ must have a vanishing eigenvalue.  Indeed,
\begin{equation}\label{uzero}
    u_0=
    \begin{pmatrix}
        nD^{-1}q\\
        nBq\\
        n
    \end{pmatrix}
\end{equation}
is a null vector of $K'$.  Related to that, the change of variables \eqref{primefic} shows that $\check{\hat b}$ decouples from the Chern-Simons terms in \eqref{KprimeLag}.  This means that, unlike the case of nonzero $p$, we cannot conclude that the low-energy theory is $\cal T'$.

One way to proceed is to add a kinetic term for $\check{\hat b}$.  Then, this gauge field is massless, and it can be dualized to a massless scalar $\phi$.

A more careful analysis of the theory does not integrate out $\Phi$ and $\hat c$.  We return to the original transition theory with all its gauge fields (without their kinetic terms) and the complex scalar $\Phi$, and discuss it in more detail.  We represent $\cal T$ as in \eqref{peelinganyonfromT} with $p=0$
\ie\label{peelinganyonfromTa}
 {\cal T} \cong {{\cal T}\boxtimes {\cal T}_{n,0}\over \bZ_n}\,,
\fe
with $ {\cal T}_{n,0}$ described by the twisted gauge field $\check {\hat b}$ and $\hat c$, and we couple $\Phi$ to $\hat c$ as above.

To better understand the IR properties of the Higgs phase, we first ignore the one-form gauging that twists $\check{\hat{b}}$, and treat it as an ordinary $\U$ gauge field. Then we will restore the quotient in \eqref{peelinganyonfromTa}.

\bigskip\bigskip\centerline{\bf Treating $\check {\hat b}$ as an ordinary dynamical $\U$ gauge field}\bigskip

Let us first ignore the fact that $\check {\hat b}$ is a twisted gauge field, and treat it as an ordinary $\U$ gauge field.  This means that we ignore the $\bZ_n$ quotient in \eqref{peelinganyonfromTa}.  Then, our transition theory factorizes into two decoupled sectors.  One of them is the original TQFT $\cal T$, and the other is $\Phi$ coupled to a $\bZ_n$ gauge theory, represented by $\U\times\U$ gauge fields $\check {\hat b}$ and $\hat c$.

The one-form symmetry of the first factor is ${\cal G}_{\cal A}^{(1)}$.  And the one-form symmetry of the second factor is $\bZ_n^{(1)}$ generated by $e^{i\int \hat c}$. (The quotient in \eqref{peelinganyonfromTa}, which we temporarily ignore, is by a diagonal $\bZ_n^{(1)}\subset {\cal G}_{\cal A}^{(1)}\times \bZ_n^{(1)}$.)

In addition, the second factor also has a $\U_\Phi^{(0)} $ symmetry, whose current and charge are
\ie\label{JPhid}
&J_\Phi={1\over n}\mathscr{J}(\Phi) = {1\over 2\pi }d\check{\hat b}\\
&Q_\Phi=\int J_\Phi\,.
\fe
Here, we used the equation of motion of $\hat c$ and the fact that we treat $\check{\hat b}$ as an ordinary $\U$ gauge field, whose fluxes are in $2\pi\bZ$.  For later use, we note that since $J_{\cal B}= \ell J_\Phi$, the symmetry generated by $Q_\Phi$ realizes
\ie\label{UPhiUB}
\U_\Phi={\U_{\cal B}\over \bZ_\ell}\,.
\fe

In the $\mu^2<0$ phase, $\Phi$ Higgses the $\bZ_n$ gauge theory of $\check{\hat b}$ and $\hat c$, and the low-energy theory includes a massless compact scalar $\phi \sim\phi+{2\pi\over n}$.  It originates from the phase of $\Phi \sim e^{i\phi}$, and the identification follows from the $\bZ_n$ gauge symmetry of this sector.  (Earlier in this subsection, we mentioned this scalar as the dual of $\check{\hat b}$.)

Restricting attention to $\phi$, the current \eqref{JPhid} of the $\U_\Phi^{(0)}$ UV symmetry can be expressed as $J_{\Phi}\sim\star d\phi$ with the normalization depending on the kinetic term of $\phi$.\footnote{ Another perspective on this gapless phase and its zero-form symmetry is the following.  Since $p=0$,  $\cal O$ of \eqref{COelld} is independent of the dual scalar $\tilde \Phi$.   In terms of the variables \eqref{primefic}, it is
\ie\label{monopolepz}
{\cal O}={\cal M}_{\check{\hat b}}\,,
\fe
i.e., it does not involve the other gauge fields.  (This is related to the fact that the coefficients in \eqref{primefic} are the same as $u_0$ \eqref{uzero}.)  In other words, in this dual description, $\cal O$ is a pure monopole operator of the massless gauge field $\check {\hat b}$.   This means that in the negative $\mu^2$ phase, the pure monopole operator $\cal O$ has a nonzero expectation value.  Since  $Q_\Phi({\cal O})=1$, its expectation value signals the spontaneous breaking of $\U_{\cal B}$ to $\bZ_\ell$, and the massless scalar $\phi$ is the Goldstone boson of this spontaneous symmetry breaking.}

We stated above that this sector has a $\bZ_n^{(1)}$ one-form symmetry, generated by $e^{i\int\hat c}$.  Let us see how it acts at low energies in this $\mu^2<0$ phase.

First, ignoring the gauge fields, the theory of $\Phi$ has no global one-form symmetry.  However, at low energies, it has a massless scalar $\phi \sim \phi+2\pi$ and an emergent $\U_{\widetilde {W}}^{(1)}$ one-form symmetry (vorticity) with the current $J_{\widetilde{W}}={1\over 2\pi} d\phi$.  Clearly, this is not an exact symmetry of the theory of $\Phi$; winding configurations, where $\phi$ winds from $0$ to $2\pi$, can unwind by going through $\Phi=0$.

Next, we add the gauge fields $\check{\hat b}$ and $\hat c$.  They represent a $\bZ_n$ gauge theory, which identifies $\Phi \sim e^{2\pi i \over n}\Phi$.  In terms of the low-energy theory, this identifies $\phi \sim \phi +{2\pi \over n}$.  Therefore, the low-energy theory now has a global $\U_W^{(1)}$ winding symmetry with the current and charges
\ie
&J_W={n\over 2\pi}  d\phi\\
&W_i=\oint_{{\cal C}_i} J_W\,,
\fe
with ${\cal C}_i$ a one-cycle.  This $\U_W^{(1)}$ is a $\bZ_n$ extension of the $\U_{\widetilde{W}}^{(1)}$ mentioned above.  Again, this is an emergent symmetry.  Only the extending $\bZ_n^{(1)}\subset \U_W^{(1)}$ generated by $e^{i\int \hat c}=e^{i\int d\phi}$ is an exact symmetry of this sector.  And therefore, only $W_i\bmod n$ are meaningful in the full theory.

The $\U_\Phi^{(0)}$ symmetry and the emergent $\U_W^{(1)}$ winding symmetry have a mixed anomaly.  Below, we will be interested in the mixed anomaly between the exact $\bZ_n^{(1)} \subset \U_W^{(1)}$ of this sector and its $\U_\Phi^{(0)}$.  A way to see it is to subject the system to a background $\bZ_n^{(1)}$ gauge field.  The latter twists $\check {\hat b}$ and makes its periods fractional, $\int {d\check{\hat b}\over 2\pi}= {t\over n}+\bZ$ with integer $t$.   Therefore, such a background $\bZ_n^{(1)}$ gauge field makes the Hilbert space realize the global $\U_\Phi^{(0)}$ symmetry projectively; all the states have $Q_\Phi ={t\over n}+\bZ$.

Consider the Hilbert space $\mathcal{H}(\Sigma_g)$ of the theory on a Riemann surface of genus $g$, $\Sigma_g$, with no insertions.  It factorizes as
\ie\label{Hbeforeg}
\mathcal{H}(\Sigma_g) = {\cal H}_{\cal T}(\Sigma_g)\otimes{\cal H}_{\Phi}(\Sigma_g) \,,
\fe
where ${\cal H}_{\cal T}(\Sigma_g)$ is the Hilbert space of the TQFT $\cal T$, and ${\cal H}_{\Phi}(\Sigma_g)$ is the Hilbert space of the decoupled theory of the $\bZ_n$ gauge theory with $\Phi$.  The states in the second factor are characterized by the conserved winding numbers $W_i\bmod n$ and the charge $Q_\Phi$.

Semiclassically, the states with nonzero $W_i\bmod n$ have higher energy than the gapless modes with $W_i\bmod n=0$.  This is related to the fact that $\U_\Phi^{(0)}$ is spontaneously broken, while $\bZ_n^{(1)}$, and even the emergent $\U_W^{(1)}$, are unbroken \cite{Gaiotto:2014kfa}.

Let us discuss these states in more detail.  Consider the Wilson line
\ie\label{calWofb}
{\cal W}_{n_{\check{\hat b}}}({\cal C})=\exp\left(in_{\check{\hat b}}\oint_{\cal C} \check {\hat b}\right)\,.
\fe
It does not generate a one-form symmetry in the theory with $\Phi$.  However, it transforms under the $\bZ_n^{(1)}$ symmetry generated by $\exp\left(i\oint\hat c\right)$ as $\exp \left({2\pi i \over n}n_{\check{\hat b}}\right)$.  This means that for  $n_{\check{\hat b}}\ne 0\bmod n$, this line is confined.  More physically, this line is the boundary of a confining string.

This can be viewed in two complementary ways.  First, place the theory on a rectangular torus with cycles ${\cal C}_1$ and ${\cal C}_2$, with sizes $L_1$ and $L_2$, and let $\cal C$ in \eqref{calWofb} run in the interior of the torus along  ${\cal C}_1$.  This means that $\phi$ winds around ${\cal C}_2$, with winding number $W_2=n_{\check{\hat{b}}}\mod n$.  Then, the energy of the state on the torus is proportional to $\int (\nabla\phi)^2 \sim {L_1\over L_2}W_2^2$.    For large $L_1\sim L_2$, such states have higher energy than the low-energy non-winding states, whose energy is of order $1\over L_1$.  Alternatively, if $\cal C$ is topologically trivial and large, its correlation function does not have a perimeter law, but it exhibits logarithmic confinement.  Below, we will deform the theory by $\cal O$, and the loop will then exhibit an area law with nonzero string tension.

\bigskip\bigskip\centerline{\bf Treating $\check {\hat b}$ as a twisted dynamical gauge field}\bigskip

Now, we restore the $\bZ_n$ quotient in \eqref{peelinganyonfromTa}. This means that we gauge the diagonal $\bZ_n^{(1)}$ generated by 
\eqref{anyoninquoK} 
\ie\label{ZngenPhi}
\exp\left(i\oint \left(q^\mathrm{T} \check c  -\hat c\right)\right)\,.
\fe
It is the product of $q$ acting on $\cal T$ and $e^{-{2\pi i\over n}W_i}$ acting on the theory of $\Phi$.

As always with gauging, we have twisted and untwisted sectors.  In the untwisted sector, we project the Hilbert space before gauging \eqref{Hbeforeg} on states that are invariant under  \eqref{ZngenPhi}. Wrapping $q$ around each one-cycle ${\cal C}_i \subset \Sigma_g$, we assign to every state in ${\cal H}_{\cal T}(\Sigma_g)$ in a basis of simultaneous eigenstates a $\bZ_n$ phase $\exp\left({2\pi i r_i\over n}\right)$ for each one-cycle ${\cal C}_i$. (Since $p=0$, the actions of $q$ around different cycles commute and can be simultaneously diagonalized.)  Then, to make this state invariant under the diagonal $\bZ_n^{(1)}$ between the two sectors, we match it with a state in ${\cal H}_{\Phi}(\Sigma_g)$  with $W_i=r_i\bmod n$.  Importantly, every state in $\cal T$ leads to infinitely many states in the full theory.  And the states with $r_i\ne 0\bmod n$ must be matched with states with $W_i\ne 0$.  They represent strings that are charged under an unbroken global $\bZ_n^{(1)}$.

We should also consider twisted sector states.  In $\cal T$, these are states that include a power of $q$ defect. As in \eqref{nonzeroRie}, in order to have such states, the genus $g$ should be large enough, and then the defect is a power of $q^{n \over \ell}$, say $q^{nt\over \ell}$.  Such a state should be matched with a twisted state from the sector of $\Phi$ with a defect $ \exp \left(-{i n t\over \ell}\int \hat c \right)$, leading to $\int {d\check{\hat b}\over 2\pi}={t\over \ell}+\bZ$ and therefore,
\ie\label{fractionalQPhi}
Q_\Phi={t\over \ell}+\bZ\qquad, \qquad Q_{\cal B}=\ell Q_\Phi=t \bmod \ell\,.
\fe
This is consistent with our statement above that twisted states in the theory of $\Phi$ have fractional $Q_\Phi$.  It is easy to see that such twisted sector states are constructed out of the untwisted states by acting on them with $({\cal O}')^t$ with $\cal O'$ of \eqref{COellp}.

Comparing this spectrum with the UV symmetry, we learn the following.
\begin{itemize}
    \item The UV ${\cal G}_{{\cal A}_q^\perp}^{(1)}$ is spontaneously broken to $\bZ_n^{(1)}\subset {\cal G}_{{\cal A}_q^\perp}^{(1)}$, which acts on low-energy strings.  (Recall that since $p=0$, $L=n$.) This $\bZ_n^{(1)}$ is enhanced to $\U_W^{(1)}$ that also acts on  strings.  The zero-energy states are  $\U_W^{(1)}$ neutral (and hence, they transform trivially under $\bZ_n^{(1)}$).  They realize the quotient ${\cal G}_{{\cal A}_q^\perp}^{(1)}\over \bZ_n$.  There is also an emergent one-form symmetry that acts on the modes of $\cal T$.
    \item The UV $\U_{\cal B}^{(0)}$ is spontaneously broken to $\bZ_\ell^{(0)}$.
\end{itemize}

The discussion of the untwisted sector above can also be phrased in terms of line operators as follows.  Consider a line operator in $\cal T$ associated with the anyon $s\in {\cal T}$.  It braids with $q$ as $e^{2\pi i n_s\over n}$. (See Section \ref{focusq}.)  To make it braid trivially with \eqref{ZngenPhi}, we multiply it by
$ {\cal W}_{n_{\check{\hat b}}} $ of \eqref{calWofb} with $n_{\check{\hat b}}=n_s$.  This creates a genuine physical line operator in the transition theory, which transforms as $e^{2\pi i n_s\over n}$ under the $\bZ_n^{(1)}$ global symmetry of that theory.

We learn that the lines in $\cal T$ that do not braid trivially with $q$ are not simply eliminated from the theory.  Instead, after multiplying them by appropriate operators, such as \eqref{calWofb}, they yield physical operators.  And then, as discussed above, these lines are confined.  Consequently, these lines do not act in the low-energy theory.

When $\cal C$ in \eqref{calWofb} is topologically trivial, ${\cal C}=\partial\cal M$ (or when it has several disconnected topologically nontrivial components bounding $\cal M$), we can use \eqref{calWofb} to define another operator, which is not a genuine line operator.\footnote{We follow the terminology of \cite{Kapustin:2014gua}.}  Instead of  adding a line operator from $\cal T$ along $\cal C$, we construct a surface operator on $\cal M$
\ie\label{linetosur}
{\cal W}_{n_{\check{\hat b}}}(\partial{\cal M})\to \widetilde {\cal W}_{n_{\check{\hat b}}}({\cal M})=&\exp\left(in_{\check{\hat b}}\int_ {\cal M} d\check {\hat b}\right)\\
=&\exp\left(2\pi in_{\check{\hat b}}\int_ {\cal M} J_\Phi\right) = \exp\left({2\pi in_{\check{\hat b}}\over \ell}\int_ {\cal M} J_{\cal B}\right)\,.
\fe
Before gauging the diagonal $\bZ_n^{(1)}$, $\check{\hat b}$ was an ordinary $\U$ gauge field and ${\cal W}_{n_{\check{\hat b}}}(\partial {\cal M})=\widetilde {\cal W}_{n_{\check{\hat b}}}({\cal M})$.  After the gauging, the left-hand side is not well-defined, and the arrow indicates how we replace it with a well-defined operator.\footnote{The reader might be puzzled that ${\cal W}_{n_{\check{\hat b}}}$ in the theory before gauging $\bZ_n^{(1)}$ can lead to different kinds of operators after the gauging.  Let us demonstrate it in a simpler example. Consider two decoupled  Ising systems, each with its own $\bZ_2$ symmetry with order parameters $\sigma_1$ and $\sigma_2$.  Then, gauge the diagonal $\bZ_2$.  Now, $\sigma_1$ and $\sigma_2$ are not gauge-invariant local operators.  (They are analogs of ${\cal W}_{n_{\check{\hat b}}}$.)  But we can still use them to form two kinds of operators/defects.  First, we have the local gauge invariant operator $\sigma_1\sigma_2$, which is analogous to  ${\cal W}_{n_{\check{\hat b}}}$  times a line in $\cal T$.  Second, we can attach $\sigma_1$ to an open symmetry line of the dual $\bZ_2$.  This is known as a twist field or a disorder operator.  It is analogous to $\widetilde{\cal W}_{n_{\check{\hat b}}}$ of \eqref{linetosur}.  And, of course, we can attach any local gauge-invariant operator to this last object.

Our case involves gauging a one-form symmetry rather than a zero-form symmetry; therefore, the dimensions of the defects/operators are increased by one relative to this simple example.}

This means that $\widetilde{\cal W}_{n_{\check{\hat b}}}$ is not a genuine line operator; it depends on additional data associated with $\cal M$.  In this form, we identify it as a variant of the $\bZ_\ell^{(0)}$ symmetry operator that acts on the open surface $\cal M$.

\subsubsection{$p= 0$ with $\cal O$ in the Lagrangian}

Now, we add the monopole operator \eqref{monopolepz} to the Lagrangian.  As for nonzero $p$, the only difference in the symmetries is that now, $\U_{\cal B}$ is explicitly broken to $\bZ_\ell$.

The effect on the low-energy theory is easy to work out.  The monopole operator gaps $\check {\hat b}$ or, equivalently, $\phi$, without affecting the other fields.  This is consistent with the fact that after breaking $\U_{\cal B}\to \bZ_\ell$, there is no massless Goldstone boson.

In more detail, the only low-energy modes are those that originated from $\cal T$.  Since the modes of $\phi$ with nonzero $W_i$ are heavier, the only states in the untwisted sector are those in the untwisted sector of ${\cal T}\over \bZ_n$.  Adding the twisted sector states, we learn that the low-energy theory is precisely ${\cal T}\over \bZ_n$.

As above, let us summarize the symmetries.
\begin{itemize}
    \item The UV ${\cal G}_{{\cal A}_q^\perp}^{(1)}$ is spontaneously broken to $\bZ_n^{(1)}$.  The unbroken $\bZ_n^{(1)}$ does not act on the low-energy theory. It acts only on massive strings.
    \item The UV $\bZ_\ell ^{(0)}$ is unbroken.  But it does act on the low-energy twisted sector states in ${\cal T}\over \bZ_n$.
\end{itemize}

Since for $p=0$ with $\cal O$ in the Lagrangian, the two phases exhibit ${\cal T}$ and $ {\cal T}'\cong {{\cal T}\over \bZ_n}$, we would like to compare this process to the process of gauging $\bZ_n^{(1)}$ as an operation on a TQFT.  (See \cite{Cheng:2026qax} for a discussion of the history of this process and its terminology.)
\begin{itemize}
    \item In the gauging picture, the original one-form symmetry ${\cal G}_{\cal A}^{(1)}$ is first reduced to  ${\cal G}_{{\cal A}_q^\perp}^{(1)}$ as a result of an ABJ anomaly between ${\cal G}_{\cal A}^{(1)}$ and $\bZ_n^{(1)}={\cal G}_{{\cal A}_q}^{(1)}$.  In the transition theory, this explicit symmetry breaking happens already at tree level in the UV.
    \item In the gauging picture, the process of removing some anyons is referred to as confinement. Importantly, the gauging picture does not exhibit confinement; the spectrum does not include confining strings.\footnote{More generally, a topological theory cannot exhibit confinement, because a nonzero string tension cannot be topological.} This is to be contrasted with the physics of the transition theory, where the $\bZ_n^{(1)} $ symmetry is unbroken, and the spectrum includes massive confining strings that are charged under $\bZ_n^{(1)}$.  These strings end on genuine line operators that are constructed using lines in $\cal T$ that braid nontrivially with $q$ (times appropriate Wilson lines \eqref{calWofb}).  We see that describing the elimination of lines in the gauging process as confining them is precisely what happens in the transition theory.
    \item In the gauging picture, there is a quotient of ${\cal G}_{{\cal A}_q^\perp}^{(1)}$ by $\bZ_n^{(1)}$.  In the transition theory, this identification happens because $\bZ_n^{(1)}$ is unbroken.  The spectrum includes massive strings that are charged under this symmetry.  But this symmetry does not act in the low-energy theory.
    \item In the gauging picture, gauging $\bZ_n^{(1)}$ leads to a quantum (or dual) $\bZ_n^{(0)}$ symmetry.  This symmetry is not present in $\cal T$, and only the quotient $\bZ_\ell^{(0)}={\bZ_n^{(0)}\over \bZ_{n\over \ell}}$ acts faithfully on ${\cal T}'\cong {{\cal T}\over \bZ_n}$.  In the transition theory, this $\bZ_\ell^{(0)}$ symmetry is an exact unbroken symmetry of the theory in all its phases.   In the phase with $\cal T$, it does not act in the IR, and in the phase with ${\cal T}'$, it acts faithfully in the IR.  This can be seen explicitly by following the description of the twisted states around \eqref{fractionalQPhi}.  Importantly, this $\bZ_\ell^{(0)}$ does not necessarily act on the torus.  Instead, as in \cite{Kobayashi:2025ykb}, it can act only on higher genus surfaces.
\end{itemize}

We end this section about the phases by summarizing in Table \ref{modelsandsymmetries} the UV symmetries of the theory with added $\cal O$ and their IR fates.

\begin{table}[ht]
\begin{align*}
\left.\begin{array}{|c|c|c|}
\hline
\mbox{\diagbox{\text{The phase}}{\text{UV symmetry}}}
&{\cal G}_{{\cal A}_q^\perp} ^{(1)}& \bZ_\ell^{(0)} \\
%&&\\
\hline
&&\\
\mu^2>0\ , \text{any}\  p& \text{Spontaneously broken} &\text{Unbroken } \\
& \text{Enhanced in the IR to}\ {\cal G}_{\cal A}^{(1)} &\text{Not acting in the IR }  \\
{\cal T}& {\cal A} \subseteq {\cal T}&\text{Unfaithful}\\
&&\\
\hline
&&\\
\mu^2<0\ , \  p\ne 0& \text{Spontaneously broken }  & \text{Unbroken} \\
 & \text{Enhanced in the IR to }\  {\cal G}_{{\cal A}'}^{(1)} \ & \text{Not acting in the IR}\ \\
 {\cal T'}\cong {{\cal T}\boxtimes{\U_{-np}} \over \bZ_n}& {\cal A}' \cong {{\cal A}\boxtimes{\U_{-np}} \over \bZ_n}\subseteq {\cal T'}\ &\text{Unfaithful}\\
&&\\
\hline
&&\\
\mu^2<0\ , \  p=0& \text{Spontaneously broken to}\  \bZ_n^{(1)}\ & \text{Unbroken} \\
 & {{\cal G}_{{\cal A}_q^\perp}^{(1)}\over \bZ_n}\ \text{enhanced in the IR to}\ {\cal G}_{\cal A'}^{(1)}& \text{Acting in the IR}\   \\
{\cal T}'\cong {{\cal T} \over \bZ_n}& {\cal A}' \subseteq {\cal T}'&\text{Faithful}\\
&&\\
\hline
\end{array}\right.
\end{align*}
\caption{The global symmetries and their fates in the various cases with $\cal O$ turned on.  The left column presents the theory (labeled by $p$), its phase (labeled by the sign of $\mu^2$), and the TQFT of the IR theory.  The top row presents the UV one-form and zero-form symmetries.}  \label{modelsandsymmetries}
\end{table}
\section{Coupling to an external $\U_A$}
\label{sec:coupling to external A}

In this section, we couple the system to an external $\U_A$ gauge field $A$, and discuss the relation between this background gauge field and the global $\U_{\cal B} $  symmetry of Section \ref{globalU1}.

Repeating the steps in Section \ref{BackgroundB}, the most general coupling of $A$, up to field redefinitions and higher order terms, is 
\ie\label{AtoTPhi}
&-A\left({1\over 2\pi}v^\mathrm{T} dc+w\mathscr{J}(\Phi)\right)\\
&v\in \bZ^N\qquad,\qquad w\in \bZ \\
&(v,w)\sim (v+D{\bf k},w-{\bf k}^\mathrm{T}DBq)\,.
\fe
(See our conventions in footnote \ref{gaugeconv}.) Unlike the discussion in Section \ref{BackgroundB}, now, we do not impose the minimality condition $t^\mathrm{T}u+z n=\ell$.  Therefore, since $\gcd(n,u^1,u^2, \cdots,u^N)=\ell$ we have 
\ie\label{Qdef}
&Q_A(\mathcal{O})=n_v+w n=Q\ell\qquad, \qquad Q\in \bZ\\
&n_v=v^\mathrm{T}u\,.
\fe
In the analogous problem in Section \ref{BackgroundB}, we imposed $Q=1$.

Continuing as in the discussion for $\cal B$, $A$ couples to the on-shell current
\ie\label{JAintermsofJB}
J_A={Q\ell\over n} \mathscr{J}(\Phi)= QJ_{\cal B}\,.
\fe

This has the following interpretation.  The integer $Q$ determines how $A$ couples to the $\U_{\cal B}$ zero-form symmetry current $J_{\cal B}$.  For $Q=0$, the current $J_A$ vanishes on-shell, and no physical operator is charged under $\U_A$.  For other values of $Q$, the $\U_A$ charge of gauge-invariant operators is $Q$ times their $\U_{\cal B}$ charge.  This is as expected, because the theory has only one faithfully acting global $\U$ symmetry, $\U_{\cal B}$.  And, on-shell, any $\U$ gauge field, such as $A$, should couple to an integer multiple of its current.

Let us review the independent parameters.  

On-shell, there is only one free parameter $Q\in \bZ$.

The off-shell coupling to $A$ involves more parameters.  One way to enumerate them is given by \eqref{AtoTPhi}.  Then, the free parameters are $v\in {\cal A}$ and $w\in \bZ$.  (We use the identification in \eqref{AtoTPhi} to let $v$ denote an anyon rather than an $N$-dimensional vector.)  Alternatively, we can say that given $Q$, we choose any $v$ satisfying \eqref{Qdef} and the other allowed $v$'s are related to it by adding $r\in {\cal A}_q^\perp$, as in the discussion around \eqref{tQPhishift}.  From this perspective, $Q$ determines how $A$ couples to the zero-form symmetry, and the freedom in $r$ labels a possible enrichment using the one-form symmetry.

We can also state it as follows.  In terms of the original TQFT $\cal T$, we identify $n_v\bmod n$ as the $\bZ_n^{(1)}$ representation of the anyon $v$ under the one-form symmetry generated by our preferred anyon $q$.  As discussed in Section \ref{focusq}, since $v\in {\cal A}\subseteq {\cal T}$, the integer $n_v$ is a multiple of $\ell$ and therefore $J_A$ is indeed an integer multiple of $J_{\cal B}$.

When $n_v=0\bmod n$, the anyon $v\in {\cal T}$ braids trivially with $q$ and then, $v\in {\cal A}_q^\perp$.  It is an element of the one-form global symmetry of the transition theory.  In that case, following \eqref{Qdef}, we can take $w=-{n_v\over n}=-v^\mathrm{T}Bq\in \bZ$, such that $Q=0$, and on-shell $J_A$ in \eqref{JAintermsofJB} vanishes.  This means that the coupling of $A$ to the transition field theory is a pure enrichment coupling.

On the other hand, when $n_v\ne 0\bmod n$, the anyon $v\in {\cal A}$ is not an element of the one-form symmetry of the transition theory ${\cal A}_q^\perp$, but the expressions above are still correct.

Let us discuss the coupling to $A$ in the low-energy theory of the two phases of the theory.  See \cite{Cheng:2026qax} for additional details.

For positive $\mu^2$, the low-energy theory is the TQFT $\cT$, and then $\U_A$ is coupled to this TQFT via enrichment. In more detail, this coupling is given by the vison $v$, and the $\U_A$ charge assignment of all anyons is completely fixed by their braiding with $v$. Explicitly, for $s\in \cT$, its $\U_A$ charge is $Q_A(s)=\sum_i\frac{n_s^iv_i}{n_i}\bmod 1$.  In particular, the $\U_A$ charge assigned to $q$ is $Q_A(q)=v^\mathsf{T}Bq={1\over n}v^\mathsf{T}u=\frac{n_v}{n}$ mod 1, as expected.  This is consistent with the fact that $\cal O$ creates the trivial anyon $q^n$ and its $\U_A$ charge $Q_A({\cal O}) = Q\ell$ is an integer.

For negative $\mu^2$ and nonzero $p$, we end up with a gapped phase with the TQFT $\cal T'$.  It is enriched by $\U_A$.  For $p=0$, we have a gapless phase associated with a nonzero $\langle{\cal O}\rangle$.  This means that if $Q\ne 0$, $\U_A$ is spontaneously broken.  And if $A$ is dynamical, it is Higgsed, as in a superconductor.  In that case, we cannot turn on $\cal O$, because that would explicitly break $\U_A$.  If, on the other hand, $Q=0$, $Q_A({\cal O})=0$, and there is no problem adding $\cal O$ to the Lagrangian and gapping the system.

\section{Conclusions and Discussion}\label{conclusions}

We presented a systematic description of the Abelian anyons in a general TQFT $\cT$ using $\U$ gauge fields with Chern-Simons terms. With this new formalism, we revisited the field theory for the proliferation of a single Abelian anyon studied in \cite{Cheng:2026qax}. This enabled us to find a more complete understanding of the theory's global symmetry and phases.

In particular, we studied the important integer $\ell$, which determines the subgroup of the theory's $\U_{\cal B}$ zero-form symmetry that does not act on local operators.  We found an explicit expression for it using only the data of the Abelian anyons $\cA$, independently of other details of $\cT$.

We analyzed in detail the IR behavior of the global symmetries in the different phases.  See Table \ref{modelsandsymmetries}. In particular, we analyzed the vortices in the $p=0$ Higgs phase where $\U_{\cal B}$ is spontaneously broken.  Interestingly, they are charged under an unbroken one-form symmetry, and hence, they bound confining strings.

In the version of the transition theory, where the global $\U_{\cal B}$ symmetry is explicitly broken to a discrete $\bZ_\ell$ symmetry, the $p=0$ theory interpolates between a phase where the low-energy theory is $\cal T$ and a phase where the low-energy theory is ${\cal T}\over \bZ_n$.  From a low-energy point of view, the map between the two phases looks like gauging a one-form symmetry.  By embedding these two phases into the non-topological transition theory, we have provided a dynamical process that implements the map between TQFTs \cite{receipe, Cheng:2026qax}.  In this framework, some of the standard aspects of this map between TQFTs, and in particular the change in the global symmetries, have a new interpretation in terms of the transition theory.  In particular, no one-form symmetry is gauged, and there is no new dual/quantum symmetry.  See Table \ref{modelsandsymmetries}.

While we focused on the transition field theory for the proliferation of a single type of Abelian anyons, it is straightforward to extend the theory to allow the proliferation of more than one type of anyons, by using multiple scalar fields coupled to distinct $q$'s. An example of such a multi-critical point is the self-dual transition in a $\bZ_N$ toric code, where both the electric and magnetic anyons proliferate. A candidate field theory similar to the approach in this work was proposed in \cite{shi_analytic_2024}. See \cite{Ji:2026yfj, Dumitrescu:2026vre} for other constructions of this field theory for $N=2$.

The transitions considered in \cite{Cheng:2026qax} and here describe the proliferation of  Abelian anyons. There are other transitions out of a TQFT in which non-Abelian anyons proliferate. Some of them can be described by Chern-Simons-Higgs theories with non-Abelian gauge groups and matter fields in higher-dimensional representations (see \cite{Cordova:2025eim, Lu:2026fid,  Garcia-Sepulveda:2026fud} for recent examples). It will be interesting to consider generalizations of the duality \eqref{TKgaugetc} that focus on non-Abelian anyons. Related ideas have been explored in \cite{Schafer-Nameki:2026hgx, Cheng:2026hgu}.

We can also entertain other transitions by modifying the dynamics of $\Phi$. If one temporarily freezes the dynamics of $\hat{c}$ and treats it as a classical background field, $\Phi$ undergoes a condensation transition in which a global $\U$ symmetry is spontaneously broken. Instead, we can change the dynamics of $\Phi$ such that the matter sector undergoes different transitions, e.g., going from a trivially gapped phase to a topological phase.  Then, we can restore the dynamics of $\hat c$ to find a new transition theory out of $\cal T$.

\section*{Acknowledgments}
We benefited from discussions with Maissam Barkeshli, Xie Chen, Clay Cordova, Dan Freed, Terry Gannon, Po-Shen Hsin, Wenjie Ji, Ho Tat Lam, John McGreevy, Gregory Moore, Seth Musser, Abhinav Prem, Marvin Qi, Brandon Rayhaun, Amir Raz, Shu-Heng Shao, Sahand Seifnashri, Senthil Todadri, Ashvin Vishwanath, and Edward Witten. MC thanks Chao-Ming Jian and Zhaoyu Han for communications about their upcoming papers.
MC is supported by NSF grant DMR-2424315.  NS is partially supported by DOE grant DE-SC0009988 and the Simons Collaboration on Ultra-Quantum Matter, which is a grant from the Simons Foundation (Grant No. 651444).

\noindent\emph{Note added:}  While the manuscript was being finalized, we became aware of two upcoming works \cite{JianTopSF, GaoTopSF} on related topics.

\appendix

\section{The transition theory based on $\hat K$ is more general than the theory based on $K$}\label{KvshatK}

Here, we show that the transition theory based on the K-matrix of \eqref{KforcalK} with $\Phi$ coupling to $f= {m}^\mathrm{T}D b+ q^\mathrm{T} c$ of \eqref{fdefinition} is less general than the theory based on $\hat K$ of \eqref{KforcalKeer}  with $\Phi$ coupling to $\hat c$.

We start with the theory based on $K$. To make it similar to the theory based on $\hat K$, we follow the discussion in Section \ref{focusqA} and introduce two more $\U$ gauge fields $\hat b$ and $\hat c$.  We couple $\Phi$ to $\hat c$ and add to the Lagrangian
\ie\label{fdefinitionbc}
{1\over 2\pi}\hat b(d\hat c -df)= {1\over 2\pi}\hat b(d\hat c - {m}^\mathrm{T}D db- q^\mathrm{T} dc) \qquad ,\qquad q,{m}\in \bZ^N\,.
\fe
So far, we have not modified the theory.  $\hat b$ acts as a Lagrange multiplier enforcing $\hat c=f$ up to a gauge transformation.

The resulting theory is very similar to the one based on $\hat K$.  The only difference is that the latter has an additional parameter, $P\in 2\bZ$, while if we start with $K$, we have only $P=0$.

Therefore, to compare the two theories, we need to look for a field redefinition in the theory with $\hat K$ that sets $P=0$.  The available redefinitions are the ones in \eqref{parameterchange}.  However, it is easy to see that, depending on $K$, $q$, and $P$, such a redefinition might not exist.

As an example of $K$, $q$, and $P$ that cannot be brought to $P=0$, consider the case
\ie
&g=\gcd(n_1,n_2,\cdots,  n_N, q^1,q^2,\cdots, q^N)\ne 1\\
&n_ip_i\in 2g\bZ\,.
\fe
Then, the redefinitions \eqref{parameterchange} cannot change $P\bmod 2g$. Therefore, in this case, the theory with $K$ cannot describe situations with $P\ne 0\bmod 2g$.  As a concrete example, take ${\cal T}={\cal A} =\U_8$ with $q=4$.  It has $n=2$, $g=4$, and $p=0\bmod 4$.  The interesting case with $p=0$ is obtained with $\hat K$ with $P=-2$, but cannot be described using $K$.

We conclude that the simpler theory with $K$ and $q$ describes most, but not all, cases. And the more complicated theory with $\hat K$ describes the most general case.

\bibliographystyle{JHEP}
\bibliography{Hall.bib}
\end{document}